\documentstyle[12pt,amsmath]{article}
\renewcommand{\theequation}{\thesection.\arabic{equation}}
\newcommand{\fract}[2]{{\textstyle\frac#1#2}}
\def\dslash{D \kern-6pt /}

\def \lwd#1{\lower2pt\hbox{$\scriptstyle #1$}}

\begin{document}
\thispagestyle{empty}
\noindent\hspace*{\fill}  FAU-TP3-00/04 \\
\noindent\hspace*{\fill}  DIAS-STP-00-08  \\
\noindent\hspace*{\fill}  MIT-CTP-2974  \\
\noindent\hspace*{\fill}  hep-th/0004200 \\  
\noindent\hspace*{\fill}  April 27, 2000\\
\rightline{\emph{with corrections in proof}}
\vspace{1in}

\begin{center}\begin{Large}\begin{bf}  Phases and Residual Gauge
Symmetries\\[1ex]
 of Higgs Models
\end{bf}\end{Large}\vspace{.75cm}

 \vspace{0.5cm} Frieder Lenz\\ Institut f\"ur Theoretische Physik III \\
Universit\"at Erlangen-N\"urnberg, Erlangen Germany\\  
\vspace{0.2cm} John W. Negele\\ Center for Theoretical Physics, MIT,
Cambridge Ma. USA\\
\vspace{0.2cm} Lochlainn O'Raifeartaigh\\  Dublin Institue for Advanced
Studies, Dublin Ireland\\  
\vspace{0.2cm} Michael Thies \\ Institut f\"ur Theoretische Physik III \\
Universit\"at Erlangen-N\"urnberg, Erlangen Germany\\
\end{center}
\vspace{1cm}

%
\begin{abstract} \noindent  After elimination of the redundant variables,
gauge theories may still exhibit  symmetries associated with the gauge
fields. The role of these {\em residual gauge symmetries} is discussed
within the Abelian Higgs model and the Georgi-Glashow model. In the
different phases of these models, these symmetries are realized
differently. The  characteristics of emergence and disappearance of the
symmetries are studied in detail and the implications for the dynamics in
Coulomb, Higgs, and confining  phases are discussed.
\end{abstract}

\newpage
\baselineskip=14pt

\section{Introduction} 
 
An appreciation of the deep and pervasive role of symmetries in
quantum mechanics and field theory has been one of the great  triumphs
of twentieth-century theoretical physics. Indeed, the connection between
symmetries and massless particles is so compelling that one's
understanding of the symmetries of a system must be fundamentally
incomplete if it cannot account for the massless excitations and particles
arising in all the phases of that system.  By this criterion, there are still
gaps in the understanding of symmetries in gauge theories.

The goal of this work is to clarify the role of \emph{residual gauge
symmetries} and their relation to the phases of gauge theories, 
thereby filling  some of these gaps.  By residual gauge symmetries we
mean symmetries associated with gauge fields that remain present after a
complete gauge fixing. In particular, these residual symmetries cannot be
generated by the Gauss law operator for one of two reasons, either
because of some geometrical property or because of some dynamical
obstruction. These residual symmetries will be studied in both the 
 Abelian Higgs model
\cite{Englert64,Higgs64,Guralnik64} 
and in the non-Abelian Georgi-Glashow model
\cite{Georgi72}. These models exhibit a rich variety of phases and
realizations of residual gauge symmetries, and  provide examples of
residual symmetries that arise from geometrical properties and from
obstructions. 

Quite generally,
electrodynamics, after elimination of redundant variables, must exhibit a
continuous symmetry that by its spontaneous breakdown produces
massless particles. Otherwise, the zero value of the photon mass would
become  accidental after elimination of the redundant variables. Indeed,
the photons can be interpreted as Goldstone bosons associated with a shift
(``displacement'') symmetry
(cf.~\cite{Ferrari71,Brandt74,Kugo,Lenz94a,Lenz94b}). In our discussion of
the Abelian Higgs model, we discuss the peculiar properties of this
symmetry  and investigate its fate in the Higgs phase. As is well known
(for example, see~\cite{ChengLi}), the Higgs phase, when formulated in
terms of the physical degrees of freedom of the unitary gauge, does not
exhibit  any continuous gauge-like symmetry nor is it a signature of the
original global phase symmetry  present; on the other hand, the system
does not contain massless particles  either. In the Abelian Higgs model,
the unitary gauge condition is optimal for describing the Higgs phase; its
choice, however, is not mandatory. For the discussion of the symmetry we
will implement the  Coulomb gauge. This is technically more involved and
the formalism is less transparent;  however, the gauge choice  is not
restricted to the description of the Higgs phase and is obviously quite
appropriate in the Coulomb phase. In this description we shall be able to
trace the disappearance of the two continuous symmetries of the Abelian
Higgs model and to understand why this disappearance is not
accompanied by the appearance of Goldstone bosons. In this discussion it
will prove useful to assume one of the space-time coordinates to be
compact; in this way  the subtle infrared properties of the model
associated with the displacements can be controlled and simple
topological interpretations of some of our results can be obtained. It will
allow us to discuss the realization of the displacement symmetry at finite
temperature and to clarify its possible violation by Debye screening. We
will  demonstrate that the requirement of invariance under
displacements imposes severe constraints on the structure of
finite-temperature effective Lagrangians. Finally, for our discussion of the
non-Abelian center symmetry
\cite{Susskind79,McLerran81,Kuti81} the introduction of a compact
space-time coordinate is indispensable.  

 The non-Abelian Higgs model, the Georgi-Glashow model \cite{Georgi72},
exhibits complementary symmetry properties. In the confining phase, the
phase with heavy adjoint Higgs scalars, the system does not exhibit any
residual symmetry nor does it contain massless particles. We
argue  it is even not quite correct to count the center symmetry which is
realized in the confining phase  as a proper symmetry;  rather we have to
consider it  as a discrete leftover of the redundancy expressed by the
non-Abelian gauge symmetry and to take it as an indication of an
incomplete gauge fixing. In the Higgs phase, on the other hand, the
physical degrees of freedom of the unitary gauge contain photons and
therefore in the transition from the confining to the Higgs phase a
displacement-like symmetry is expected to be present in the Hamiltonian
or Lagrangian describing the Higgs phase. Also, the rather mysterious
center symmetry has to be broken by the Higgs vacuum. Otherwise static
fundamental charges should have a confining interaction, which is
certainly not expected  in the presence of physical massive
and massless vector particles. Within the unitary gauge we will describe
the mechanisms by which a symmetry absent in the confining phase can
emerge and specify  how the center symmetry, a discrete part of the
gauge symmetry, is spontaneously broken. As in the Abelian case, a
representation of the dynamics in the confining and Higgs phase within
one common gauge choice can be attained provided the gauge condition
does not involve the Higgs field. Unlike the Coulomb phase of the Abelian
Higgs model, the confining phase of Yang-Mills theories,  as well as  its
relation to the Higgs phase, is understood only poorly. We will attempt to
shed some light on the connection between the Higgs and confining phases.
To this end, we will derive an effective Coulomb-gauge Lagrangian for the
description of the Higgs phase and with the help of ``gauge-invariant'' 
variables extend this discussion to other gauge conditions. This will allow
us to study qualitatively the fading of possible  mechanisms  of
confinement in the transition to the Higgs phase.

\section{Abelian Higgs Model}\setcounter{equation}{0}
\subsection{Residual Gauge Symmetries} In this section we will discuss
residual gauge symmetries and their realization in the two phases of the
Abelian Higgs model. In standard notation, the Lagrangian is given by
\begin{equation}
  \label{Hila}
  {\cal L}\left[A,\Phi\right]=-\fract{1}{4}
F^{\mu\nu}F_{\mu\nu}+(\partial_{\mu}+ieA_{\mu})
\Phi(\partial^{\mu}-ieA^{\mu})\Phi^{\star}-V(|\Phi|).
\end{equation}
 The Higgs potential $V(|\Phi|)$ and the coupling constant $e$ determine
 the phases of this model. As emphasized in the Introduction, we also
 assume the system to be of finite extent $L$ in at least one of the
 spatial directions (the 3-direction). In conventional treatments, one
normally derives the properties of these phases by choosing
 gauges that simplify the dynamics in the relevant regime. In the
 weak coupling limit ($e \ll 1$ -- see below) of the Abelian Higgs
 model, any gauge choice that constrains the gauge field such as the
Lorentz  or Coulomb gauge is appropriate. On the other hand, the
description of the strong coupling limit, i.e., of the Higgs phase, simplifies
in the ``unitary'' gauge, which constrains the Higgs field and thereby
displays directly the particle content of this phase.
 In this work, however,  we shall perform an analysis within a single
gauge choice. Although this
 necessarily will complicate matters technically, it  will enable us to
 investigate the fate of the symmetries in the two phases
 and will lead to a unified dynamical description.

 In the definition of
the generating functional
\begin{equation} Z[J,k]= \int d[A, \Phi]\,\delta [f({\bf A},\Phi)]
e^{iS\left[A,\Phi\right]}e^{i\int d^{4} x\,
\left(J^{\mu}A_{\mu}+k\Phi\right)}
\label{gg1}
\end{equation}
 we assume the gauge condition
\begin{equation}
  \label{gc} f({\bf A},\Phi)=0
\end{equation} 
does not affect the time component of $A$. In a large part of this section
we will choose the Coulomb gauge. We also will use in part of our
discussion the polar representation for the Higgs field
$$\Phi = \rho e^{i\varphi}.$$ After implementation of the Coulomb gauge,
residual symmetries are still present.  Clearly the system described by
the above gauge fixed functional is invariant under two global, continuous
symmetries, the
    {\em global (rigid) phase changes}
\begin{equation}
\Phi(x) \rightarrow e^{i\alpha}\Phi(x) \qquad(\varphi(x)
\rightarrow \varphi(x) +\alpha)  
\label{gpc}
\end{equation}
 and  {\em displacements} \cite{Lenz94b}
\begin{equation} {\bf A}(x) \rightarrow {\bf A}(x)+\frac{1}{e}\,{\bf
d}\qquad\Phi(x) \rightarrow e^{i{\bf d}\cdot {\bf x}}\Phi(x)
\qquad \bigl(\varphi(x) \rightarrow \varphi(x) +{\bf d}\cdot {\bf
x}\bigr).
\label{dps}
\end{equation} These transformations do not change the action nor do
they  change the gauge condition. Strictly speaking, displacements are not
continuous transformations, since periodicity in the 3-direction enforces
quantization of the 3-component of the displacement vector
$$
d_{3} = \frac{2\pi n}{L} .
$$ 
It is obvious that a gauge choice that
constrains the gauge fields does not affect the global phase symmetry of
the theory. The displacement symmetry is a direct consequence of the
equations of motion. With the help of current conservation, Maxwell's
equations can be written as
\begin{equation}
  \label{maxw}
\partial_{\mu} (F^{\mu\nu}-ej^{\mu}x^{\nu})=0 .
\end{equation}  In this way, the components of Maxwell's displacement
vector
$$
{\bf D} = \int d^{3}x\,\left({\bf E}+e{\bf x}j^{0}\right) 
$$ 
are identified
as the generators of displacements. We note that along a compact direction
$$ 
\int d^{3}x\,E^{i} \neq -\int d^{3}x\,x^{i}\mathop{\mathrm div}{\bf E}
$$
 and thus
displacements cannot be generated by the Gauss\ law operator
$G=\mbox{div}\,{\bf E}-ej^{0}$. Therefore, after gauge fixing,
displacements are present as residual symmetries and  the zero modes,
$\int  d^{3}x {\bf A}(x)$, are gauge invariant.
 In a gauge-fixed
formulation, the interpretation of displacements as residual symmetry
transformations becomes manifest. In the Coulomb gauge, with the
transverse fields as dynamical degrees of freedom, the symmetry
transformation can be softly modulated
\begin{equation} {\bf A}(x) \rightarrow {\bf
A}(x)+\frac{1}{2e}\mathop{\mathrm rot}\left[{\bf d}({\bf x})\times {\bf
x}\right]\qquad \quad\Phi(x) \rightarrow e^{i{\bf d}({\bf x})\cdot {\bf
x}}\Phi(x) .
\label{dpsl}
\end{equation} 
This transformation is clearly not a gauge transformation;
the gauge condition is respected by construction and it reduces in the
infinite wavelength limit to the symmetry transformation (\ref{dps}).
Thus for large wavelength modulations the restoring forces generated
when applying this transformation to the ground state will be weak and
the excited states therefore of low energy. Unlike other symmetry
transformations, the displacement symmetry has the peculiar property
that in a certain sense, it always gives rise to gapless excitations (or if the
3-space is compact,  excitations with energies
$\sim 1/L$.) The existence of an operator whose commutator with ${\bf
D}$ has a nonvanishing expectation value is guaranteed kinematically by
the canonical commutation relation
\begin{equation}
  \label{bdds}
  \left[D_{k}, \frac{1}{V} \int  d^{3}x\,A_{l}\right] = i \delta_{kl}.
\end{equation}  
Thus Goldstone bosons associated with this symmetry
must exist.  The connection to a possibly nontrivial structure of the
vacuum is more subtle. In Appendix I the physical content and some
general issues concerning spontaneous symmetry breakdown are
discussed. We will address the issue of the structure of the vacuum in the
context of the Georgi-Glashow model. Related to  the kinematical nature of
the appearance of the photons as Goldstone bosons  is the fact  that the
Ward identities associated with the displacements and the rigid phase
transformations actually imply the Maxwell equations (\ref{maxw})
(cf.~Appendix II). 

The presence of two continuous global symmetries has to be expected.
The dynamics has not yet been specified and therefore the above
gauge-fixed generating functional could, for example, describe decoupled
radiation and matter. In this case, the theory would contain massless
photons enforced by the displacement symmetry and scalars that as well
may  be massless as a result of the spontaneous breakdown of the 
rigid phase symmetry.  On the other hand, as is well known, there are no
massless excitations in the Higgs phase, nor does the Higgs phase exhibit
any symmetries
\cite{ChengLi,ORaifeartaigh79}. The disappearance  of the residual gauge
symmetries in the Higgs phase without emergence of massless particles is
the main subject of the following  two sections.

\subsection{The Higgs Phase}  In this section we derive in Coulomb-like 
gauges the effective Lagrangian describing the dynamics in the Higgs
phase. On the basis of this development we will give a complete account
of the symmetries present in the original form of the Lagrangian. To this
end we have to integrate out the constrained variables contained in the
gauge fixed Lagrangian, the time component of the gauge field. The action
is quadratic in the field
$A_{0}$. The integration can be carried out exactly resulting in an
expression for the generating functional which neither contains
integration over redundant nor over constrained variables. A
straightforward calculation yields for general gauge conditions of the
form (\ref{gc}) and with vanishing time component of the source $J$
\begin{equation} Z[J,k]= \int d[{\bf A}, \Phi]\, \delta [f({\bf A}, \Phi)]
e^{iS_{\mathrm eff}\left[{\bf A},\Phi\right]}e^{i\int d^{4}\, x \left(-{\bf
J}\cdot{\bf A}+k\Phi\right)}
\label{gg2}
\end{equation} 
with
$$
S_{\mathrm eff} = \int d^4 x\,\left( {\cal L}_{\mathrm kin}+{\cal
L}_{\mathrm pot}\right)
-\fract{1}{2}\mathop{\mathrm tr}\left\{\ln{\Gamma}\right\}
$$
\begin{eqnarray} 
{\cal L}_{\mathrm pot} & = & -\fract{1}{4}
F^{ij}F_{ij}+(\partial_{i}+ieA_{i})\Phi(\partial^{i}-ieA^{i})\Phi^{\star}
-V(|\Phi|)\\
 {\cal L}_{\mathrm kin}& =
&(e\rho^{2}\partial_{0}\varphi-\fract{1}{2}\partial_{i}
\partial_{0}A^{i})\,
\frac{2}{\Gamma}\,(e\rho^{2}\partial_{0}
\varphi-\fract{1}{2}\partial_{i}
\partial_{0}A^{i})+\rho^{2}(\partial_{0}\varphi)^{2}+
(\partial_{0}\rho)^{2}+\fract{1}{2}(\partial_{0}A^{i})^2\nonumber 
\label{lsp}
\end{eqnarray} 
and
\begin{equation}
  \label{dtm}
\Gamma = \Delta - 2e^{2}\Phi\Phi^{\star} .
\end{equation} In Coulomb gauge, 
$$
f({\bf A}, \Phi) = \mathop{\mathrm div} {\bf A}
$$
the expression for $ {\cal L}_{\mathrm kin}$
simplifies
\begin{equation}
  \label{ltgc}
 {\cal L}_{\mathrm kin} =
\rho^{2}\partial_{0}\varphi\,\frac{1}{\Gamma}\,\Delta
\,\partial_{0}\varphi +
    (\partial_{0}\rho)^{2}+\fract{1}{2}(\partial_{0}{\bf A})^{2} .
\end{equation} 
Up to this point, we have not yet used any property that
is specific to the Higgs phase.  As in the standard treatment, we now
assume  that the Higgs phase emerges if the self-interaction
$V$
 generates a large expectation value $\rho_{0}$ of the Higgs field and that
it becomes meaningful to use the polar representation. We decompose the
modulus of the Higgs field into this vacuum expectation value and
a fluctuating piece
$$
\rho(x) = \rho_{0}+\frac{1}{\sqrt{2}}\sigma(x)
$$ 
 and keep only terms
quadratic in the fields $\sigma,\varphi$, and ${\bf A}$. In the resulting
expression (cf.~Eq.~(\ref{lsp}))
$${\cal L}_{\mathrm pot}\approx  -\fract{1}{4} F^{ij}F_{ij} -\fract{1}{2}
(\mbox{\boldmath$\nabla$}\sigma)^{2}-\rho^{2}_{0}
 (\mbox{\boldmath$\nabla$}\varphi+e{\bf A})^{2}
$$ 
we single out the
zero modes of the fields $\mbox{\boldmath$\nabla$}\varphi$ and  {\bf A}
\begin{equation}\mbox{\boldmath$\nabla$}\varphi+e{\bf
  A}=(\mbox{\boldmath$\nabla$}\varphi+e{\bf A})_{0}+(
\mbox{\boldmath$\nabla$}\varphi+e{\bf A})^{\prime},\quad\mbox{with}
  \int d^{3}x{\bf A}^{\prime} = \int
  d^{3}x\mbox{\boldmath$\nabla$}\varphi^{\prime} = 0,
\label{zm1}
\end{equation} and note that ${\bf A}^{\prime}$ and $\varphi^{\prime}$
are invariant under displacements. We obtain
$$S_{\mathrm eff} \approx \int d^4\, x {\cal
    L}_{\mathrm eff}
$$ 
 with
\begin{equation}
  \label{hlg}
 {\cal L}_{\mathrm eff}= -\fract{1}{2} {\bf A}^{\prime}(\Box +m^2)
 {\bf A}^{\prime} -
\fract{1}{2}\chi(\Box+m^2) \chi - \fract{1}{2}\sigma\Box
\sigma -\fract{1}{4}V^{\prime\prime}(\rho_{0})\sigma^{2}+{\cal L}_{0}.
\end{equation} The  mass of the vector mesons is
$$
m^{2}=2e^{2}\rho_{0}^{2}
$$ 
 and their longitudinal component is given by
\begin{equation}
  \label{chi}
 \chi = \rho_{0}\left(\frac{-2\Delta}{-\Delta
+m^{2}}\right)^{\fract{1}{2}}\varphi^{\prime}.
\end{equation}
 The zero-mode dynamics is described by
\begin{equation}
  \label{zm} {\cal L}_{0} =\fract{1}{2} ({\bf
\dot{A}}_{0})^{2}-\rho^{2}_{0}
 (\mbox{\boldmath$\nabla$}\varphi+e{\bf A})_{0}^{2} .
 \end{equation} No time derivative of
$(\mbox{\boldmath$\nabla$}\varphi)_{0}$ appears so it is not
a dynamical variable. As far as the  3-component of this vector is
concerned, this time independence is a consequence of the periodicity of
the scalar field, which  obviously requires
$$ 
\int_{0}^{L} dz\, \partial_{3} \varphi = 2n\pi 
$$ 
  and therefore does not
permit continuous changes of the 3-component of
$(\mbox{\boldmath$\nabla$}\varphi)_{0}$. 
Since $(\mbox{\boldmath$\nabla$}\varphi)_{0}$ is essentially an
irrelevant constant we may absorb it in ${\bf A}_0$ and write
\begin{equation}
 {\bf A}_{0}+\frac{1}{e}(\mbox{\boldmath$\nabla$}\varphi)_{0}
\rightarrow {\bf A}_{0} .
\label{zma}
\end{equation} 
Thus $ {\cal L}_{0}$  provides the zero modes for the
massive photon field. 

 It is interesting how, in Coulomb gauge, the standard properties of the
Higgs phase emerge. Like in the unitary gauge, the mass of the transverse
gauge fields is generated by their coupling to the Higgs condensate. The
massive longitudinal degrees of freedom appear due to  a dynamically
modified kinetic energy of the phase $\varphi$ of the Higgs field
(cf.~Eq.~(\ref{ltgc})). Rewriting the kinetic energy in  canonical form, Eq.~(\ref{chi}) makes the transformation of the massless compact variables
$\varphi$ into the massive noncompact fields $\chi$ manifest. In the
Coulomb gauge, the crucial quantity that distinguishes Coulomb  and
Higgs phase is the operator $\Gamma$ of Eq.~(\ref{dtm}). In the Higgs
phase, the relevant field configurations must generate a gap in the
spectrum of
$\Gamma$. In this case,    the fluctuations of $\rho^{2}$ around a
nonvanishing expectation value may be neglected. In turn, the transition
from the Higgs to the Coulomb phase has to be accompanied by
appearance and possible condensation of zero modes of~$\Gamma$. This is
the mechanism   of formation of a  condensate of vortices
\cite{Abrikosov57,Nielsen73}in the unitary gauge. In unitary gauge, the
appearance of a vortex signifies a failure of the gauge condition -- the
gauge transformation that eliminates the phase of the Higgs field is ill
defined at those points where the Higgs field vanishes.
The association of singularities with the zeroes of the Higgs field
is thus a gauge artifact and  has no physical significance.

 We now resume our discussion of the residual symmetries in the Abelian
Higgs model. Our construction shows that in the space of redefined fields
(cf.~Eqs.~(\ref{chi}), (\ref{zm1}) and (\ref{zma})) global phase changes
(Eq.~(\ref{gpc})) and displacements (Eq.~(\ref{dps}))
\begin{equation}
  \label{sym}
  \chi(x) \rightarrow \chi(x) \qquad {\bf A}^{\prime}(x)\rightarrow {\bf
A}^{\prime}(x)
\end{equation}
 are reduced to identity transformations.  Thus, in the Higgs phase  the
Lagrangian does not exhibit any symmetries, as one might have expected. 

 The mechanism that makes the rigid phase symmetry disappear is
displayed by Eq.~(\ref{ltgc}). In the Higgs phase, it is not $\varphi$ that
is a dynamical variable, rather it is
$\sqrt{-\Delta}\varphi$. Thus the global phase of the Higgs field is not
dynamical. The physics behind the redefinition of the phase variable is
easily understood in the Coulomb gauge. The Coulomb interaction of the
charged scalar field in the Higgs phase
$$ 
 e^{2}\int d^3x\,  d^3x^{\prime}\frac{j^{0}(x)j^{0}(x^{\prime})}{|{\bf
x}-{\bf x}^{\prime}|}\sim e^{2}\rho_{0}^{4}\int d^{3}x\,  d^{3}x
^{\prime}\,\frac{\partial_{0}\varphi(x)\partial_{0}\varphi(x^{\prime})}{|{\bf
x}-{\bf x}^{\prime}|}
$$ 
    gives rise to a nonlocal renormalization of the
kinetic energy which is accounted for by the field redefinition in Eq.~(\ref{chi}) and, by its long-range nature, prevents  the emergence of
Goldstone bosons. We also observe that in the transition from Coulomb to
Higgs phase, the compact $\varphi$ degree of freedom changes into a
noncompact one. 

 The disappearance of the displacement symmetry in the Higgs phase is
most easily understood by enclosing the system in a finite volume and
imposing boundary conditions. Otherwise, due to the linear ${\bf
  x}$-dependence of the displacements, certain integrations by parts
necessary for deriving the effective Lagrangian are ill defined. With
compact space, the absence of displacements in the Higgs phase has a
simple topological interpretation. Whenever the gauge field is coupled to
matter, the shift in the gauge field is accompanied by a change in the
winding number associated with the phase of the matter field. As in the
case of the $\theta$-vacuum, in order to form a state that is symmetric
under displacements, states with different winding numbers have to be
superimposed. Similarly, to generate Goldstone bosons in the broken
phase, the possibility for space-dependent long wavelength displacements
accompanied by corresponding changes in the winding number must
exist. This is not the case in the Higgs phase. Under continuous changes, a
field configuration with a certain winding number
$\Phi_{n}$ can only be connected to a field configuration with different
winding number
$\Phi_{n ^{\prime}}$, if these fields vanish somewhere in space. This is
effectively ruled out by the assumption of a condensate large enough to
make the effect of fluctuations negligible  to leading order. Thus in the
Higgs phase, the winding number is frozen, i.e., the quantity
$(\mbox{\boldmath$\nabla$}\varphi)_{0}$ is not a dynamical variable
and therefore the displacement symmetry is not present. In other words,
sectors with different winding numbers are separated by barriers of
infinite energy, associated with the discontinuous changes in the field
configurations. The same topological property that freezes the winding
number also prevents the emergence of Goldstone bosons  in this
symmetry breakdown. The system therefore does not  offer in the Higgs
phase the possibility for displacements. When  the strength of
the condensate is continually decreasing, the density of strings of zeroes
in 3-space increases and makes it  progressively easier for the system to
change winding numbers.  Eventually, in the perturbative phase,
we assume implicitly that the topological constraint is absent as in the
free Maxwell theory and quantum fluctuations of the Higgs field around
zero field no longer provide any obstacle for change in winding number.
More formally, in the perturbative limit the relevant mapping is
$x^{3}\in S^{1} \rightarrow \Phi\in C$ rather than   the topologically
nontrivial $x^{3}\in S^{1}
\rightarrow \varphi\in S^{1}$ relevant for the Higgs limit. 

   In the Higgs phase,  the constant gauge field absorbs part of the phase of
the Higgs field (cf.~Eq.~(\ref{zma})) and thereby turns into the
displacement invariant variable $ {\bf A}_{0}^{\prime}$. Thus, emergence
of Goldstone bosons is avoided  because in the Higgs phase  the system 
contains only dynamical variables that are invariant under
displacements. In a description where we introduce from the beginning
the phase of the Higgs field as a dynamical variable, this resolution of
the symmetry issues is easily seen. In the Higgs phase, the transformation
to the  {\em gauge-invariant variables}
 $$
B_{\mu}= A_{\mu}+\frac{1}{e}\partial_{\mu}\varphi,\quad\rho
$$ 
 is
well defined; in terms of these variables, the generating functional
(\ref{gg1}) is given by
\begin{equation} 
Z = \int d[B]\prod_{x}\rho(x)\,d\rho(x)\int
d[\varphi]\,\delta [f(B-\frac{1}{e}\,\partial\, \varphi ,\rho
e^{i\varphi})]e^{i
\int d^{4} x\,  {\cal L}\left[B,\rho\right]}
\label{ggin}
\end{equation}    
with the Lagrangian
\begin{equation}
  \label{Hilu}
  {\cal L}\left[B,\rho\right]=-\fract{1}{4}
F^{\mu\nu}[B]F_{\mu\nu}[B]+e^{2}\rho^{2}
B_{\mu}B^{\mu}+\partial_{\mu}\rho\,\partial^{\mu}\rho-V(\rho).
\end{equation} 
The integration over $\varphi$,  which also includes a
summation over
 the winding number, can be used to satisfy the gauge condition. The
 remaining integration over the zero mode of~$\varphi$ factorizes in~$Z$.
Thus the generating functional is completely determined by the
 fields~$B$ and~$\rho$. These variables are invariant under both
 global phase changes and displacements. Thus, we find once more that
 in the Higgs phase these two symmetry transformations become the
 identity; in other words,  the corresponding symmetries are
 implemented trivially. We note that in this way a conflict between
 the ``realization'' of the displacement symmetry and its breakdown
 enforced by the canonical commutation relation (\ref{bdds}) is avoided.

\subsection{Scalar QED at Finite Temperature} 
In this concluding section
on the Abelian Higgs model, we briefly
 discuss the possible effects of thermal fluctuations  on the
 realization of the displacement symmetry. To this end, we consider the
Abelian Higgs model in the Coulomb phase and
 assume the Higgs potential in Eq.(\ref{Hila}) to be given by a  mass term
$$
V(|\Phi|)= m^{2}\Phi\Phi^{\star}.
$$ 
 At sufficiently high
 temperatures, the photon is coupled to a finite density of charged
 particles and  may thereby acquire a mass. Such a mechanism is
apparently  operative leading to Debye screening. However, in
perturbation theory no ``magnetic'' mass is generated. The presence of the
Debye mass is not compatible with the displacement symmetry. We study
this issue by  integrating  out the scalar field in the generating functional
(\ref{gg1}), and drop its source term
 \begin{equation}
   \label{ft1}
 Z[J]= \int d[A]\,\delta [f(A)] e^{-S_{\mathrm eff}[A]}e^{i\int d^{4}\, x
J^{\mu}A_{\mu}}.   
 \end{equation} 
The gauge condition has been assumed to be independent
of the scalar field. The generating functional is written in Euclidean
space and, following the usual convention, the compact direction is
denoted in this section as the 0-direction. The effective action~is
\begin{equation}
  \label{ft2}
 S_{\mathrm eff} = \fract{1}{4}\int d^{4}\, x F_{\mu\nu} F_{\mu\nu} +
 \mathop{\mathrm tr}\ln \left[-D_{\mu}D_{\mu}+m^2\right].  
\end{equation}
 To lowest order in a  perturbative expansion, the
determinant is given by tadpole and virtual pair diagrams, which for
vanishing photon momentum cancel for ``spacelike'' gauge fields and
yield  at high temperatures  ($T=1/L$)  the well-known Debye mass
(\cite{KAPU},\cite{LEBE})
$$
m_{D}^{2} = \fract{1}{3} e^{2}T^{2} 
$$ 
 associated with $A_{0}$. Inclusion
of this lowest order result leads to an effective Lagrangian that is not
invariant under displacements. On the other hand, the `tr\,
log'--contribution to the action is invariant. For an
eigenfunction~$\chi$
$$
\left[-D_{\mu}D_{\mu}+m^2\right]\chi = \lambda \chi 
$$ 
 the
transformed eigenfunction
$$
\tilde{\chi}=e^{id_{\mu}x_{\mu}}\chi 
$$ 
 satisfies
$$
\left[-\tilde{D}_{\mu}\tilde{D}_{\mu}+m^2\right]\tilde{\chi} = \lambda
\tilde{\chi}
$$ 
 with the displacement-transformed covariant derivatives
$$
\tilde{D}_{\mu} =D_{\mu}-i\,d_{\mu} .
$$ 
  Thus the spectrum $
\left\{\lambda\right\}$ is invariant under displacements.  This
invariance indeed excludes, for $\mu=1,2,3 $, any dependence of the
effective action on the zero modes of the gauge fields 
$$
a_{\mu} = \frac{1}{V} \int_{V} d^{4}x\, A_{\mu}(x) 
$$ 
 and therefore no
magnetic mass can be generated. The invariance  under the discrete
displacements in 0-direction
$$
d_{0}= \frac{2\pi n}{L}
$$ 
  also rules out a purely quadratic mass term for $a_{0}$. However, it is
compatible with a periodic dependence of $S_{\mathrm eff}$ 
$$
S_{\mathrm eff}[a_{0}]= S_{\mathrm eff}[a_{0}+\frac{2\pi}{eL}]
$$ 
which as we will show below can also be locally quadratic for $a_0$ near
a multiple of $2\pi/eL$. 
 Since displacements
contain  the inverse of the electric charge, a perturbative evaluation
violates the displacement symmetry. Therefore we will evaluate the
determinant in the effective potential approximation; i.e., we take into
account only zero-momentum photons 
$$
 \mathop{\mathrm tr}\ln \left[-D_{\mu}D_{\mu}+m^2\right]\approx  \mathop{\mathrm tr}\ln
\left[-(\partial_{\mu}+iea_{\mu})^{2}+m^2\right] = V_{\mathrm eff}(a).
$$ 
 To
calculate $V_{\mathrm eff}$ we add terms that are independent of the
gauge-field zero modes and obtain
\begin{eqnarray*} V_{\mathrm eff}(a) &=& \mbox{const.}+\frac{1}{L}
\sum_{n=-\infty}^{\infty}\int
\frac{d^{3}k}{(2\pi)^{3}}
\left[\ln\left(({\bf k}+e{\bf
a})^{2}+\kappa_{n}^{2}(a_0)\right)-\ln\left({\bf
k}^{2}+\kappa_{n}^{2}(0)\right)\right]\\ &=&
\mbox{const.}+\frac{1}{L}\sum_{n=-\infty}^{\infty}\int^{\kappa_{n}^{2}(a_0)}_{\kappa_{n}^{2}(0)}
dM^{2}\int \frac{d^{3}k}{(2\pi)^{3}}\frac{1}{{\bf
    k}^2+M^{2}}=-\frac{1}{6\pi
L}\sum_{n=-\infty}^{\infty}|\kappa_{n}(a_0)|^{3}   
\end{eqnarray*}
 with
$$
\kappa_{n}^{2}(a_0) = \Bigl(\frac{2\pi n}{L}+ea_{0}\Bigr)^{2}+m^{2}.
$$ 
  The
momentum integral has been evaluated in dimensional
regularization. The dependence on the spacelike components of the gauge
field has disappeared in the shift of the integration variables. The final
sum is calculated with
$\zeta$-function regularization~(cf.~\cite{ELIZ96})
$$
V_{\mathrm eff}(a) = -\frac{m^2}{\pi^{2}
  L^{2}}\sum_{n=1}^{\infty}\frac{1}{n^{2}}\cos(n e L a_{0})K_{2}(nmL).
$$   
  The displacement symmetry is explicitly  preserved in the
   evaluation of the effective potential. 
  Formally we may define a Debye mass by adding and subtracting the
quadratic
  term of the expansion of the cosine
$$
m_{D}^{2}= \frac{e^{2}m^{2}}{\pi^{2}}\sum_{n=1}^{\infty}K_{2}(nmL).
$$ 
The displacement symmetry makes, however, self-interactions of $a_{0}$
beyond the mass term necessary.    In the high-temperature limit, the
$n$-sum can be performed and, up to a constant, the effective potential is
given by 
$$
V_{\mathrm eff}(a)=
\frac{2\pi^{2}}{3L^{4}}(\alpha^{2}-2\alpha^{3}+\alpha^{4})\qquad 0\leq
\alpha\leq 1,\quad mL \ll 1 
$$ 
 where 
$$
\alpha = \frac{eLa_{0}}{2\pi}.
$$ 
We observe both that the effective potential is periodic, preserving
displacement symmetry at finite temperature, and that it is locally
quadratic for~$a_0$ in the vicinity of zero with curvature corresponding
to the standard value of~$m_0$. Thus, 
 while invariance of the finite-temperature effective action under displacements rules out a magnetic
mass of the photons,  the presence of a mass in the time component of the
gauge field as part of a periodic self-interaction is compatible with the
displacement symmetry. 
In addition, if we relabel the axes and reinterpret the system as being at
zero temperature with $x_3$ periodic, in the limit of large~$L$ we expect
displacement symmetry to manifest itself by excitations with energies
$\sim1/L$.

At sufficiently high temperature we expect the preceding arguments to
apply to the Abelian Higgs model. Therefore, in addition to the restoration
of the rigid-phase symmetry in the phase transition from the Higgs to the
Coulomb phase at finite temperature, we also expect that the
displacement symmetry will reappear accompanied by gapless
excitations.  

\section{Georgi-Glashow Model}\setcounter{equation}{0}
\subsection{Symmetries} In this  section we present a general discussion
of the symmetries of the Georgi-Glashow model
\cite{Georgi72} with emphasis on the center symmetry
\cite{Susskind79,McLerran81,Kuti81}.
 The Georgi-Glashow Lagrangian describes the $SU(2)$ Yang-Mills theory
coupled to an adjoint scalar
\begin{equation}
  \label{gegl}
 {\cal L}
\left[A,\Phi\right]=-\fract{1}{4}F_{\mu\nu}F^{\mu\nu}+\fract{1}{2}D_{\mu}\Phi
D^{\mu}\Phi -V(|\Phi|)
\end{equation} with the covariant derivative of the Higgs field
$$ 
D_{\mu}\Phi = \partial_{\mu}\Phi-gA_{\mu}\times \Phi.
$$ 
 We have
written the Higgs field and the field strength tensor as  vectors in color
space.  For the following discussion, it is convenient to represent both
gauge and Higgs fields in a spherical color basis and we will  refer to
those objects pointing in the color 3 direction as neutral, those
perpendicular as charged. At this point this definition is formal. Only later
in the process of gauge fixing will the meaning of the color directions 
be specified. The charged components of gauge and Higgs fields are
defined as
\begin{equation}
  \label{gf12a}
  A^{\pm}_{\mu}\,  =
  \frac{1}{\sqrt{2}}(A_{\mu}^{1}\mp i A_{\mu}^{2})
\end{equation}
\begin{equation}
  \label{gf12b}
  \Phi^{\pm} =
  \frac{1}{\sqrt{2}}(\Phi^{1}\mp\,i\Phi^{2})
\end{equation} 
The neutral component of the field strength is  
\begin{equation}
  \label{gf18}
  F_{\mu\nu}^{3} = \partial_{\mu}A_{\nu}^{3}- \partial_{\nu}A_{\mu}^{3}-
ig(A^{-}_{\mu}A^{+}_{\nu}-A^{-}_{\nu}A^{+}_{\mu})
\end{equation}
 and the charged ones are
\begin{equation}
  \label{gf19}
  F_{\mu\nu}^{\pm}=(\partial_{\mu} \pm ig A_{\mu}^{3})A^{\pm}_{\nu}-
(\partial_{\nu} \pm ig A_{\nu}^{3})A^{\pm}_{\mu} .
\end{equation} The kinetic part of the Higgs field can be similarly
rewritten and the following form of the Lagrangian of the Georgi-Glashow
model is obtained
$$
{\cal L} ={\cal L}_{YM}+{\cal L}_{H}
$$ 
 with
\begin{equation}
  \label{gf20a}  {\cal L}_{YM} =  -\fract{1}{4} F^{3\mu\nu}
F_{\mu\nu}^{3}- \fract{1}{2} F^{+\,\mu\nu} F_{\mu\nu}^{-}
\end{equation}
\begin{eqnarray}
  \label{gf20b} {\cal L}_{H}&=&\fract{1}{2}\left[\partial_{\mu}
\Phi^{3}+ig(\Phi^{-}A^{+}_{\mu}-A^{-}_{\mu}\Phi^{+})\right]
\left[\partial^{\mu}\Phi^{3}+ig(\Phi^{-}A^{+\,\mu}-A^{-\,
    \mu}\Phi^{+})\right]
\\&& {}+D_{\mu}^{3\dagger}\Phi^{-}D^{3\,\mu}\Phi^{+}
+
ig\Phi^{3}\left[A^{-\,\mu}D_{\mu}^{3}\Phi^{+}-A^{+\,\mu}D_{\mu}^{3\,
\dagger}\Phi^{-}\right] 
 + g^{2}A^{-}_{\mu}
A^{+\,\mu}(\Phi^{3})^{2}-V(|\Phi|)  \nonumber
\end{eqnarray}
 where
$$
 D_{\mu}^{3}= \partial_{\mu}+ ig A_{\mu}^{3} .
$$ 

\paragraph{Center
Symmetry} As above, we  assume that one  of the space-time directions,
the 3-direction, is compact. In a non-Abelian gauge theory we can
associate a loop integral of gauge fields with this compact direction, the
Polyakov loop,
\begin{equation} P_{3} \left(x_{\perp}\right) =  P\exp\Bigl\{ig
\int_{0}^{L} dz\,  A_{3}
\left(x\right)  \Bigr\}.
\label{FE3}
\end{equation}
 The trace of
$P_{3}$  can serve as order parameter for the phases of Yang-Mills
theories \cite{McLerran81,Svetitsky86,Lenz98}. The expectation value and
correlation functions of this variable are related to the self-energy of a
single static quark  and the interaction energy of two static quarks
respectively and therefore distinguish the high-temperature
gluon plasma from the low-temperature confining phase. Under gauge
transformations $U(x)$, $P_{3}\left(x_{\perp}\right)$ transforms as
\begin{equation} P_{3}(x_{\perp}) \rightarrow
U\left(x_{\perp},L\right)P_{3}(x_{\perp})
 U^{\dagger}\left(x_{\perp},0\right)\ .
\label{I31}
\end{equation} The coordinates  $x=(x_{\perp},0)$ and
$x=(x_{\perp},L)$ describe identical points, and we require the periodicity
properties imposed on the field strengths not to change under gauge
transformation. This is achieved if
$U$  satisfies
\begin{equation}
  \label{cs1}
 U\left(x_{\perp},L\right) = c_{U}\cdot U\left(x_{\perp},0\right)
\end{equation} with $c_{U}$ being an element of the center of the group.
Thus gauge transformations can be classified according to the value of
$c_{U}$ ($\pm 1$ in SU(2)). Therefore under gauge transformations
\begin{equation}
\mathop{\mathrm tr}\bigl(P_{3}(x_{\perp})\bigr) \rightarrow
\mathop{\mathrm tr}\bigl(c_{U} P_{3}(x_{\perp})\bigr)
\stackrel{SU(2)}{=} \pm \mathop{\mathrm
tr}\bigl(P_{3}(x_{\perp})\bigr).
\label{I31b}
\end{equation}
 A simple example of a class of  SU(2) gauge
transformations $u_{n}$ with $c=\pm 1$ is
\begin{equation} u_{n}(x) = e^{i \mbox{\boldmath$\hat{\scriptstyle\psi}
$}(x)\cdot
\mbox{\boldmath$\scriptstyle\tau $}n\pi x_{3}/L}, \qquad
c_{u_{n}}=(-1)^{n} \label{cs1a}
\end{equation}
 with the arbitrary and, in general, space-time dependent
unit vector
$\mbox{\boldmath$ \hat{\scriptstyle\psi}$}$ in color space. The
transformed gauge and Higgs fields are
\begin{eqnarray}
  \label{cs1b} A_{\mu} &\rightarrow& u_n \Bigl( A_{\mu}- \frac{i}{g}
\partial_{\mu}
\Bigr) u_n^{\dagger}  \qquad
  \Phi \ \rightarrow \ u_{n}\Phi u_{n}^{\dagger}
 \\ u_n \partial_{\mu}
u_n^{\dagger} & = & i\left[ \left(1-\cos \Bigl(\frac{2\pi n
x^3}{L}\Bigr)\right)
\mbox{\boldmath$ \hat{\psi}$}\times\partial_{\mu}
\mbox{\boldmath$ \hat{\psi}$} -\sin \Bigl(\frac{2\pi n x^3}{L}\Bigr)
\partial_{\mu}\mbox{\boldmath$ \hat{\psi}$} -\frac{2 \pi
n}{L}\mbox{\boldmath$
\hat{\psi}$}\delta_{\mu 3}\right] \frac{\mbox{\boldmath$\tau$}}{2}
\nonumber
\end{eqnarray} 
These symmetry transformations  are reminiscent of the
displacements of the $U(1)$ theory. There are, however, important
differences related to the different topological properties of Abelian and
non-Abelian theories in the presence of a compact direction. Unlike
displacements that cannot be gauged away, since $\Pi_{1}(U(1))= Z$, the
symmetry transformations in Eq.~(\ref{cs1b}) can be continuously
connected to the identity for  even~$n$, since $\Pi_{1}(SU(2))= Z_{2}$. An
example of such a continuous deformation of these transformations into
the identity has been given in \cite{Griesshammer93}
\begin{eqnarray}
  \label{grie}
  u_{2n}(x,t) & = &e^{-(it\pi/2)\mbox{\boldmath$\hat{\scriptstyle\chi}
$}_{1}\cdot\mbox{\boldmath$\scriptstyle\tau $}}e^{(it\pi/2) \left\{
\mbox{\boldmath$\hat{\scriptstyle\chi} $}_{1}\cdot
\mbox{\boldmath$\scriptstyle\tau $}\cos 2n\pi x^{3}/L +
\mbox{\boldmath$\hat{\scriptstyle\chi} $}_{2}\cdot
\mbox{\boldmath$\scriptstyle\tau $}\sin 2n\pi
x^{3}/L\right\}}\nonumber\\
 u_{2n}(x,0) &=& 1 ,\qquad  u_{2n}(x,1) = u_{2n}(x),\qquad  u_{2n}(0,t)=
 u_{2n}(L,t)=1
\end{eqnarray} where the  unit-vectors  $\mbox{\boldmath$
\hat{\scriptstyle\psi}$},\mbox{\boldmath$
  \hat{\scriptstyle\chi}$}_{1,2}$ form a right-handed, orthogonal basis. In
order
  to single out the topologically nontrivial piece of the above
  transformation, we define center reflections by
\begin{equation} Z_{k} =  i   e^{i\pi \mbox{\boldmath$
\hat{\scriptstyle\psi}\scriptstyle_{\perp} $}\cdot
\mbox{\boldmath$\scriptstyle\tau$}/2} e^{i(2k+1)\pi
\mbox{\boldmath$\hat{\scriptstyle\psi} $}
\cdot
\mbox{\boldmath$\scriptstyle\tau $}x^{3}/L}
\quad\mbox{with}\quad\mbox{\boldmath$\hat{\psi}$}\cdot
\mbox{\boldmath$\hat{\psi}_{\perp}$} =0
 \label{cs1c}
\end{equation} where $k$ is some fixed integer. These transformations
are reflections and change the sign of the Polyakov loop
\begin{equation}
  \label{cr}
  Z^{2}_{k}=1,\quad c_{Z_{k}}=-1  .
\end{equation} Center reflections can be used to generate any other
gauge transformation changing the sign of $\mathop{\mathrm tr}(P_{3})$ by
multiplication with a strictly periodic ($c=1$) but otherwise arbitrary
gauge transformation. The decomposition of $SU(2)$ gauge
transformations into two classes according to $c=\pm 1$ implies a
decomposition of each gauge orbit $\cal O$, into suborbits
 $\cal O_{\pm}$, which are characterized by the sign of the Polyakov loop
at some fixed $x_{\perp}^{0}$
\begin{equation}
  \label{cs8a}
 A(x)\, \in \, {\cal O}_{\pm} ,\qquad \mbox{if } \pm
\mathop{\mathrm tr}\bigl(P_{3} (x_{\perp}^{0})\bigr)\ge 0 .
 \end{equation}  
Thus, strictly speaking, the trace of the Polyakov loop is
not a gauge-invariant quantity. Only $|\mathop{\mathrm
tr}(P_{3}(x_{\perp})|$ is invariant under all of the gauge transformations.
Furthermore, the spontaneous breakdown of the center symmetry in
Yang-Mills theory as it supposedly happens at small extension or high
temperature is a breakdown of the underlying gauge symmetry. It
implies that the wave functional describing such a state is different for
gauge field configurations that belong to $\cal O_{+}$ and $\cal O_{-}$
respectively, and  that therefore are  connected by gauge transformations
such as
$u_{2n+1}$ in Eq.~(\ref{cs1a}). These symmetry considerations apply
equally well for the Georgi-Glashow model. Coupling of the Yang-Mills
field to matter in the adjoint representation does not affect the center
symmetry unlike coupling to fundamental fermions. In the latter case,
center symmetry transformations change the boundary condition of fields
carrying fundamental charges. Thus realization of the center symmetry
should be equally relevant for the phases of the Georgi-Glashow model.

These considerations are also of relevance for understanding the
 structure of gauge theories after (partially) fixing the gauge. Whenever
gauge fixing is carried out exactly and with the help of strictly periodic
gauge-fixing transformations ($\Omega, c_{\Omega}=1$)
 the resulting formalism must contain the center symmetry
\begin{equation}
\mathop{\mathrm tr}\bigl(P_{3}(x_{\perp})\bigr) \rightarrow 
-\mathop{\mathrm tr}\bigl(P_{3}(x_{\perp})\bigr) 
\label{I31a}
\end{equation}  as residual gauge symmetry, in other words, each gauge
orbit is represented by two gauge-field configurations.

\subsection{Residual Gauge Symmetries in Unitary Gauge} 
We now consider the role of the residual symmetries of the
Georgi-Glashow model in the unitary gauge.  The unitary gauge condition
is
\begin{equation}
  \label{ug}
  \Phi = \Phi^{a}(x) \frac{\tau^{a}}{2}=
  \rho(x) \frac{\tau^{3}}{2} .
\end{equation} The gauge condition does not affect the gauge fields.
Therefore the Yang-Mills piece of the Lagrangian remains unchanged and
the contribution of the Higgs field  simplifies to
\begin{equation}
  \label{laug}
  {\cal L}_{H} =
\fract{1}{2}\partial_{\mu}\rho\partial^{\mu}\rho +g^{2}\rho^{ \,
2}A^{-}_{\mu} A^{+\,\mu}-V(|\rho|).
\end{equation}
 By this gauge fixing,  the color 3-direction is identified with
the direction of the Higgs field. The symmetry transformations
$U_{n}$ after gauge fixing are obtained using~Eq.~(\ref{cs1a}) with
$$
\mbox{\boldmath$\hat{\psi}$}= \hat{\Phi}.
$$ 
 With this choice, the
unit-vector $\mbox{\boldmath$\hat{\psi}$}$ becomes
space-time independent after transforming to the unitary gauge and
$$u_{n} =e^{in \pi x^{3}\tau^{3}/L} .$$  We thus consider the following
transformations
\begin{eqnarray}U_{n}= u_{2n} &:&  \quad A_{\mu}^{3}\left(x\right)
 \rightarrow  A_{\mu}^{3}\left(x\right)-\frac{4n\pi}{gL}\delta_{\mu,3}
\qquad  A^{+}_{\mu}\left(x\right)  \rightarrow  e^{4in\pi
  x^{3}/L}A^{+}_{\mu}\left(x\right)\nonumber\\ &  &  \quad
F_{\mu\nu}^{3}\left(x\right)
 \rightarrow  F_{\mu\nu}^{3}\left(x\right)\qquad \hskip 1.55cm
F^{+}_{\mu\nu}\left(x\right)  \rightarrow  e^{4in\pi
  x^{3}/L}F^{+}_{\mu\nu}\left(x\right)\nonumber \\[2ex]
 & & \quad
\,\hskip 0.25cm \rho\left(x\right)\,\,
 \rightarrow \, \rho\left(x\right) .\quad \quad \hskip 1.7cm
\end{eqnarray}
With an appropriate choice of
$\mbox{\boldmath$\hat{\psi}$}$,  center reflections $Z_{k}$
(Eq.~(\ref{cr}))  become
$$
Z_{k}=ie^{i \pi\tau^{1}/2}e^{i(2k+1) \pi x^{3}\tau^{3}/L}
 $$
and transform
the relevant variables as
\begin{eqnarray} Z_{k} &:& \quad  A_{\mu}^{3}\left(x\right)
 \rightarrow
-A_{\mu}^{3}\left(x\right)-\frac{2\pi(2k+1)}{gL}\delta_{\mu,3} \quad
\quad \hskip 0.3cm A^{+}_{\mu}\left(x\right)  \rightarrow e^{2i(2k+1)\pi
  x^{3}/L}A^{-}_{\mu}\left(x\right)\nonumber \\
 & & \quad  F_{\mu\nu}^{3}\left(x\right)
 \rightarrow  - F_{\mu\nu}^{3}\left(x\right)\quad \hskip 3.30cm
F_{\mu\nu}^{+}\left(x\right)  \rightarrow
 e^{2i(2k+1)\pi
  x^{3}/L}F_{\mu\nu}^{-}\left(x\right)\nonumber\\[2ex]
 & & \quad
\,\,\hskip 0.25cm \rho\left(x\right)\,
\, \rightarrow  -\,\rho\left(x\right).
\end{eqnarray}  With the above choice of the
$\mbox{\boldmath$\hat{\psi}_{\perp}
  $}$-dependent term, we actually have included in $Z$  a charge
conjugation\footnote{For a proper definition of C in the Higgs phase one
should include a reflection of $\rho$ at the origin, which leaves ${\cal
L}_{H}$ invariant;  the following discussion will not be affected by such a
redefinition.}  
$$Z_{k}=Cu_{2k+1}$$
\begin{eqnarray} C &:&  \quad A_{\mu}^{3}\left(x\right)
 \rightarrow  -A_{\mu}^{3}\left(x\right) \quad \quad \hskip 0.3cm
A^{+}_{\mu}\left(x\right)  \rightarrow
A^{-}_{\mu}\left(x\right)\nonumber \\
 & & \quad \qquad F_{\mu\nu}^{3}\left(x\right)
 \rightarrow  - F_{\mu\nu}^{3}\left(x\right) \quad
F_{\mu\nu}^{+}\left(x\right)  \rightarrow
F_{\mu\nu}^{-}\left(x\right)\nonumber\\ & &  \qquad \qquad \,\,\hskip
0.25cm \rho\left(x\right)\,
\, \rightarrow  -\rho\left(x\right)  .
\end{eqnarray} 

We note that these three classes of transformations leave the gauge-fixed
Lagrangian invariant
$$ 
U_{n}, \, Z_{k},\, C : \quad {\cal L}_{YM}+{\cal L}_{H} \rightarrow {\cal
L}_{YM}+{\cal L}_{H} 
$$
However, they cannot be simultaneously
implemented, since
$$
 \left[U_{n},Z_{k}\right] \neq 0\qquad   \left[U_{n},C\right] \neq 0
\qquad \left[Z_{k},C\right] \neq 0  .
$$
 In our analysis of the symmetries
we first address the displacements. As argued above, unlike
displacements in the Abelian theory, there is no topological obstruction to
a change in winding number. This is also true after gauge fixing. 
However, after gauge
fixing,  the deformations of the fields in Eq.~(\ref{grie}) are no longer 
occurring along equipotential lines. Rather, the necessary rotation of  the
gauge field in color space introduces terms of the following form:
$$
g^{2}\rho^{ \, 2}A^{-}_{\mu} A^{+\,\mu}\rightarrow g^{2}\rho^{ \,
2}A^{-}_{\mu} A^{+\,\mu} + 2g^{2}\rho^{ \, 2}A^{3}_{\mu}
A^{3\,\mu}\sin^{2}t\pi\,\sin^{2}\frac{n\pi x^{3}}{L}+\cdots
$$
 By
construction, along this path the winding of the phase of the charged
gauge-field component changes with identical values of the potential
energy at initial and final points $t=0,1$. However, although the values of
the action corresponding to these initial and final field configurations are
the same, they are  separated by potential barriers whose height
increases with the condensate. Hence, ultimately  changes in 
$A^{3\,\mu}$, and therefore in the winding number, by such deformations
involving all of the color components of the gauge field are prevented
dynamically. 
This mechanism effectively prevents the displacements from being
generated by the Gauss law operator and is an example of the second
form of residual gauge symmetry mentioned in the Introduction.
 In the Higgs limit, the system
behaves like an Abelian model and like the Abelian model in the weak
coupling limit, it possesses the possibility of changing the winding in the
phase of the charged gauge fields at their zeroes. Thus, for a sufficiently
large condensate and concomitantly large masses of the vector particles, a
physical displacement symmetry appears that,  as in weakly coupled
QED,  in turn requires the existence of Goldstone bosons. 

We now discuss the discrete symmetries, the center reflection, and the
charge conjugation. With the displacement symmetry broken by the
mechanism described in the first section the variable
$$ 2g\int_{0}^{L}A_{3}^{3}\, dz = \chi $$ assumes a fixed value, which by
a suitable redefinition of the gauge fields can be shifted to the interval
$$0 \le \chi < 2\pi .$$ In general, both discrete symmetries are broken. 
The vacuum is invariant under charge conjugation only  if $ \chi= 0$,  and
it is center symmetric only if $\chi=\pi$. The classical field configurations
of lowest potential energy (cf.~Eqs.~(\ref{gf19},\ref{laug}))
$$\rho =  \rho_{\rm min}, \quad A^{\pm}_{i} =0 ,\quad A_{i}^{3} =
\mbox{const.} \, $$ do not single out a particular value of
$A_{3}^{3}$. 

Since classical considerations cannot determine $A_{3}^{3}$, we consider
the quantum mechanical ground state energy, which is
affected by the presence of ``Aharonov-Bohm'' fluxes as described by a
nonvanishing constant gauge field. To calculate this quantity, we formally
eliminate
$\chi$ by redefining the phase of the charged vector-particles
$$\hat{A}^{\pm}_{\mu} = e^{\pm i\chi x^{3}/L}\,A^{\pm}_{\mu}$$ at the
expense of introducing modified boundary conditions
$$\hat{A}^{\pm}_{\mu}(x^{3}=L)= e^{\pm
  i\chi}\hat{A}^{\pm}_{\mu}(x^{3}=0).$$ The ground state energy density
corresponding to these boundary conditions has been calculated
 in \cite{Eletsky98} with the result that the contribution of the charged
vector-particles of
  mass $M$ is
  given by
\begin{eqnarray}
\varepsilon_{A^{\pm}} & = & -\frac{3}{\pi^2 L^4}\left[
\Bigl( \frac{1}{2 \epsilon} + \fract{3}{4} - \fract{1}{2}
\gamma                  + \ln (\sqrt{4 \pi} \mu/M) \Bigr) (ML/2)^4\right.
\nonumber
\\                                                                    & &\left. \qquad{} + (ML)^2
\sum_{n=1}^{\infty} \frac{1}{n^2} \cos (n\chi)              K_2(nML) \right]
\
\label{m5}
\end{eqnarray}       with the standard parameters $\epsilon,\mu$
appearing in dimensional regularization. Thus, the vacuum state that
is even under charge conjugation ($\chi = 0$) is favored energetically at
high temperatures where $ K_2(nML)$ is nonnegligible. However,
the suppression of nonvanishing values of
$\chi$ is exponentially small for $M\,L \gg 1$ and therefore  irrelevant at
temperatures much smaller than the vector particle mass.

To explore the low-temperature regime, we take the  approach commonly
used in studying symmetry breaking of investigating the response of
the system to an arbitrarily weak external perturbation and determining
the value of $\chi$ that leads to the lowest energy.  In the present case, we
consider the response to a pair of color charges, which can be expressed in
terms of  Polyakov loops and will therefore be directly related to
center symmetry.  For massive charged
vector-particles,  the correlation function of two Polyakov loops at large
distances is dominated by the contribution of the massless neutral fields
$A^{3}_{\mu}$ and we therefore can write
$$\langle
0|\,T(\mathop{\mathrm tr}P_{3}(x_{\perp})\mathop{\mathrm
tr}P_{3}(y_{\perp})\,|0\rangle
 \,\approx  \,  \langle 0|\,T(\mathop{\mathrm tr}e^{igL a(x_{\perp})\tau^{3}/2}
\mathop{\mathrm tr}e^{igL a(y_{\perp})\tau^{3}/2})\,|0\rangle $$ with
$$ a(x_{\perp}) = \frac{1}{L}  \int_{0}^{L} dz \, A_{3}^{3}
\left(x\right) .$$ In the Fock-space decomposition of the ground state we
treat the  zero-momentum state separately. Here we assume a
spontaneous  breakdown of the displacement symmetry with the vacuum
in  a field eigenstate $|\chi\rangle$
$$ \Bigl(\frac{1}{V_{\perp}} \int_{V_{\perp}} d^{2} x
\,gLa(x_{\perp})\Bigr) |\chi\rangle=\chi |\chi\rangle .$$ This choice is
suggested by the vanishing of the kinetic energy of these zero-modes in
the thermodynamic limit (cf.~Appendix I).The Fock-states corresponding to
finite momentum photons are  particle number eigenstates and their
contribution  can be treated in the standard way. We first calculate the
expectation value
\begin{eqnarray}
\langle 0_{\chi^{\prime}}| \mathop{\mathrm tr}\bigl(e^{igL a(x_{\perp})\tau^{3}/2}
\bigr)  |0_{\chi}\rangle &=& 2\delta(\chi^{\prime} - \chi)\cos \frac{\chi}
{2} \langle 0|\,e^{igL a^{\prime}(x_{\perp})/2} \, |0\rangle  \nonumber \\ &
= & 2\delta(\chi^{\prime} - \chi) \cos \frac{\chi}{2} e^{-\frac{g^{2}}{16
\pi} L\Lambda /2}
\label{WL13}
\end{eqnarray}
 where the matrix-elements of
$$  a^{\prime}(x_{\perp})= a(x_{\perp})-\frac{\chi}{gL}$$ are those of the
corresponding Maxwell theory and have been calculated by replacing the
sum over (transverse) momenta by integrals. The divergence in this
integral has been regulated with heat kernel regularization (regulator
$\Lambda$). Obviously, the $\delta$-function in the above expression can
be avoided by using a superposition of field eigenstates centered around
a specific value. In the evaluation of the correlation function we perform
a Euclidean rotation  ($x_{\perp}^{2} \rightarrow r^{2}$) and
obtain for the interaction energy of two static charges
\begin{eqnarray}
 e^{-LV(r)} & = & \langle 0_{\chi^{\prime}}|\mathop{\mathrm tr}\bigl( e^{igL
a(x_{\perp})\tau^{3}/2}\bigr)\mathop{\mathrm tr} \bigl(e^{\pm igL
    a(y_{\perp})\tau^{3}/2}\bigr)\,|0_{\chi}\rangle   \nonumber\\ &=&
2\delta(\chi^{\prime}  - \chi)\left[\langle 0| e^{igL
    a^{\prime}(x_{\perp})/2}e^{- igL a^{\prime}(y_{\perp})/2}|0\rangle
+\cos{\chi}\langle 0| e^{igL a^{\prime}(x_{\perp})/2}e^{ igL
    a^{\prime}(y_{\perp})/2}|0\rangle \right]
\nonumber\\ & = &
 2\delta(\chi^{\prime} - \chi) e^{-\frac{g^{2}}{16 \pi} L\Lambda
   }\Bigl[e^{\frac{g^2}{16\pi } \frac{L}{r}} +
    \cos{\chi}\,e^{-\frac{g^2}{16\pi } \frac{L}{r}}\Bigr] .
\label{WL17}
\end{eqnarray} 
Thus, for all distances, $r$,  the minimum potential $V(r)$ occurs for the
maximum of $\cos{\chi}$, favoring $\chi$ = 0. 
Although the
strength of the ``condensate'' of Aharonov-Bohm flux, $a(x_{\perp})$,  is of
the order
$L^{-1}$, this constant gauge field has a nonnegligible effect when
integrating $j_{3}A^{3}$ over an $x^3$-independent current. For large
separation ($r\gg L$), the interaction energy  of static charges behaves as
$1/r$ unless $\chi$ assumes the center symmetric value
$\pi$ or the charge conjugation symmetric value $\chi=0$. In the
center symmetric case, the interaction energy increases logarithmically
with distance, providing a hint of confinement. For the charge-conjugation
symmetric case, the potential decreases like
$r^{-2}$. We note that the modification of the ground
state energy by the static charges is  not
suppressed exponentially like the Casimir energy (Eq.~(\ref{m5})). 

Our calculations thus suggest that there should be no change in the
realization of the center symmetry and therefore no discontinuity in the
Polyakov loop as a function of temperature throughout the Higgs phase for
sufficiently large values of the condensate. With increasing temperature,
i.e., decreasing $L$, the Casimir energy increasingly favors, independent of
an external disturbance,  the charge symmetric point
(cf.~Eq.~(\ref{m5})) that was already preferred at low temperature.

\subsection{Coulomb-gauge Effective Lagrangian} In this section we
derive in Coulomb gauge the effective Lagrangian for the Georgi-Glashow
model. We proceed  as in the Abelian Higgs model and
first integrate out the nondynamical $A_{0}$ variables.  In this first step
the gauge condition only has to be assumed to be independent of
$A_{0}$ but can remain unspecified otherwise. We represent the Higgs
field by its modulus $\rho$ and the unit vector $\hat{\Phi}$ defining its
orientation. The resulting effective action, including the Faddeev-Popov
operator $\Delta_{FP}$ is given by
$$S_{\mathrm eff} = \int d^4 x\,\left( {\cal L}_{\mathrm kin}+{\cal
    L}_{sp}\right)+\mathop{\mathrm tr}\left\{\ln \Delta_{FP}\right\}
-\fract{1}{2}\mathop{\mathrm tr}\left\{\ln{\Gamma}\right\} .$$ with the  two
contributions to the effective Lagrangian
\begin{eqnarray*} {\cal L}_{\mathrm pot} & = & -\fract{1}{4}
F^{ij}F_{ij}+\fract{1}{2}\rho
 \Delta\rho -V(\rho) - \fract{1}{2}  \rho^{2}\hat{\Phi}\,
\tilde{\Gamma}\,\hat{\Phi}\\
 {\cal L}_{\mathrm kin}& = & -\fract{1}{2}(D_{i}\partial_{0}A^{i}+
g\rho^{2}\partial_{0}\hat{\Phi}\times\hat{\Phi})\frac{1}
{\Gamma}(D_{i}\partial_{0}A^{i}+g\rho^{2}\partial_{0}
\hat{\Phi}\times\hat{\Phi})\\ & &\quad{} + \fract{1}{2} \rho^{2}(\partial_{0}
\hat{\Phi})^{2}+\fract{1}{2}(\partial_{0}\rho)^{2}+\fract{1}{2}
(\partial_{0}A^{i})^2
\end{eqnarray*} with
$$\Gamma^{ab}= \tilde{\Gamma}^{ab}+
g^{2}\rho^{2}(\delta^{ab}-\hat{\Phi}^{a}\hat{\Phi}^{b})$$ and the gauge
covariant Laplacian
$$\tilde{\Gamma}^{ab}= -\delta^{ab}\Delta +g\epsilon^{abc}
(A^{ic}\partial_{i} + \partial_{i}A^{ic}) - g^{2}(\delta^{ab} A_{i}^{c}A^{i\,c} -
A_{i}^{a}A^{i\,b}) .$$ We simplify these expressions by assuming that the
Higgs field develops a large expectation value. We decompose the
modulus of the Higgs field
$$\rho(x)=\rho_{0}+\sigma(x) ,$$ and keep only terms  quadratic in the
fields $A_{i}, \partial_{\mu}\hat{\Phi},
$ and $\sigma$. Because of the constraint $\hat{\Phi}^2 = 1$, at this point
we cannot disregard terms involving higher powers of $\hat{\Phi}$. 
However, we note that terms involving derivatives of $\hat{\Phi}$ may be
truncated at quadratic order, with the result 
\begin{eqnarray}
  \label{ltp2}{\cal L}_{\mathrm kin} &\approx& 
 - \fract{1}{2}\bigl(\partial_{0}\partial_{i}A^{i}+
g\rho^{2}_{0}\partial_{0}\hat{\Phi}\times\hat{\Phi}\bigr)\,
\frac{1}{-\Delta +g^{2}\rho^{2}_{0}(1-\hat{\Phi}\otimes
\hat{\Phi})}\,\bigl(\partial_{0}\partial_{j}A^{j} +g
\rho^{2}_{0}\partial_{0}\hat{\Phi}\times\hat{\Phi}\bigr)
\nonumber \\ &&\quad{}+ \fract{1}{2} \rho_{0}^{2}(\partial_{0}
\hat{\Phi})^{2}+\fract{1}{2}(\partial_{0}\sigma)^{2}
+\fract{1}{2}(\partial_{0}A^{i})^2\nonumber\\[2ex]
 {\cal L}_{\mathrm pot}
&\approx& -\fract{1}{4} (\partial_{i}A_{j}-
\partial_{j}A_{i})(\partial^{i}A^{j}- \partial^{j}A^{i}) +\fract{1}{2}\sigma
 \Delta\sigma -\fract{1}{2} V^{\prime\prime}(\rho_{0})
\sigma^{2}\nonumber \\&&\quad{}+ \fract{1}{2}  \rho^{2}_{0} \Bigl[
  \hat{\Phi}\Delta\hat{\Phi}+2g(\partial_{i}\hat{\Phi}\times
  \hat{\Phi})\, A^{i}-g^{2}\left(A^{i}-\hat{\Phi}(\hat{\Phi}
A^{i})\right)^{2}\Bigr].
\end{eqnarray} This representation in terms of the unit vectors
$\hat{\Phi}$ is well defined in the Higgs phase with its large expectation
value of
$\Phi$.  Generation of a mass for the components of the transverse vector
  fields orthogonal to the Higgs field is explicit in the last term of
${\cal L}_{\mathrm pot}$. The appearance of the massive longitudinal
  components is more subtle. The above expression displays the special
roles of the unitary gauge and the Coulomb gauge. In these gauges, the
longitudinal components of the gauge fields and the
  orientation of the Higgs field do not mix to leading order. In unitary
  gauge the 2  massive longitudinal components are described by the fields
$$
\hat{\Phi}^{a}=\delta_{a,3}:\quad\tilde{\chi}^{a} =
g\rho_{0}\Bigl(\frac{1}{-\Delta(-\Delta
    +g^{2}\rho_{0}^{2})}\Bigr)^{\fract{1}{2}}\mbox{div}{\bf A}^{a},
\quad a=1,2 
$$
and no time derivative of $\mbox{div}{\bf A}^{3}$ is
present, reflecting the incompleteness of the unitary gauge condition. In
Coulomb gauge
 $$
\mbox{div}{\bf A}^{a} = 0, \quad a=1,2,3 
$$ 
we have to separate
massive from massless excitations. To this end we introduce two
vector-fields, an isoscalar ($c^{i}$) and a constrained isovector ($C^{i\,a}$),
by representing the gauge fields as
\begin{equation}
  \label{cv}
 A^{i} =  c^{i} \hat{\Phi}+ C^{i}\times \hat{\Phi}\,,\quad
C^{i}\cdot\hat{\Phi} = 0.
\end{equation} 
In making the harmonic approximation, we again drop terms that are
generated when  differential operators act on the unit vectors
$\hat{\Phi}$. Hence, we write
$$(\partial_{0}A^{a,i})^2 \approx (\partial_{0}c^{a,i})^2 +
  (\partial_{0}C^{a,i})^2, $$ and correspondingly  simplify the
Coulomb gauge condition
$$\delta(\partial_{i}A^{i})
\approx\delta(\hat{\Phi}\partial_{i}
c^{i}-\hat{\Phi}\times\partial_{i}C^{i}).$$ We replace the functional
integral over the 3 constrained vector-fields
$C$ by two ``Cartesian'' vector fields.   With interactions neglected, the
Faddeev-Popov determinant becomes trivial. Finally to achieve a
canonical form for the kinetic terms of the Higgs field unit vectors we
define
\begin{equation}
  \label{chi2}
 \chi = \rho_{0}\Bigl(\frac{-\Delta}{-\Delta
+g^{2}\rho_{0}^{2}}\Bigr)^{\fract{1}{2}}\hat{\Phi}
\end{equation} and with $g\rho_{0} \rightarrow \infty$ the constraint on
$\hat{\Phi}$ yields for sufficiently small momenta two Cartesian fields
$$
D[\Phi_{1},\Phi_{2},\Phi_{3}]\,
\delta\Bigl(\sum_{i=1,3}\Phi_{i}^{2}\Bigr)\propto
D[\chi_{1},\chi_{2},\chi_{3}]\,
\,\delta
\Bigl(\sum_{i=1,3}\chi_{i}\Delta\chi_{i}+g^{2}\rho_{0}^{2}\Bigr)\approx
D[\chi_{1},\chi_{2}].
  $$ 
This series of small amplitude approximations results in the final
expression for the generating functional
\begin{equation}
  \label{Zq} Z[J] = \int D[c,C^{1},C^{2},\chi^{1},\chi^{2},\sigma]\,
\delta(\partial_{i}c^{i})\prod_{a=1,2}\delta(\partial_{i}C^{i
  \,a})e^{i\int d^{4}x {\cal L}_{\mathrm eff} +j_{c}c+..}
\end{equation} with the Coulomb-gauge Lagrangian
 \begin{eqnarray}
  \label{gglg}
 {\cal L}_{\mathrm eff}&=& -\fract{1}{2}\, c_{i}\,\Box\, c_{i}
-\fract{1}{2}\,\sum_{a=1,2}\left[C_{i}^{a}\,\left(\Box+g^{2}\rho^{2}_{0}\right)
C_{i}^{a}+\chi^{a}\,\left(\Box+g^{2}\rho^{2}_{0}\right)  \chi^{a}
\right]\nonumber\\
&&\quad{} - \fract{1}{2}\,\sigma\,\Box\,
\sigma - \fract{1}{2}V^{\prime\prime}(\rho_{0})\sigma^{2}.
\end{eqnarray} As in the Abelian case, the small amplitude approximation
is essential for the derivation of the Coulomb-gauge fixed Lagrangian and
is justified a posteriori by the mass generation. In both cases  the massive
vector-particles receive their longitudinal components from the compact
variables specifying the orientation of the Higgs fields in the internal
space ($U(1)$ and $SU(2)$ respectively). This can only happen since in
the  presence of the mass gap the kinetic energy is modified and thereby 
the compact $\hat{\Phi}$ are transformed into noncompact variables. 

\subsection{Gauge-conditions in Higgs and Confining Phases}
 An efficient description of the Higgs phase can be obtained with the
 help of gauge invariant variables.  The starting point is the partition
function
\begin{equation} Z[J,k]= \int d[A, \Phi]\delta [f(A,\Phi)]\Delta_{FP}[A,\Phi]
e^{iS\left[A,\Phi\right]}e^{i\int d^{4} x \left(J^{\mu}A_{\mu}+k\Phi\right)}
\label{ggl1p}
\end{equation} with a general gauge condition
\begin{equation}
  \label{gfgg} f({A},\Phi)=0
\end{equation} and $S$ is the action of the Georgi-Glashow Lagrangian
(\ref{gegl}). We assume that the modulus of the Higgs field has a large
expectation value and parametrize the Higgs field as
\begin{equation}
  \label{diag}
  \Phi = \fract{1}{2} \rho \,U\tau^{3}\, U^{\dagger}.
\end{equation} The unitary matrix $U$ diagonalizes the Higgs field and
can be parametrized, e.g., by
\begin{equation}
  \label{para}
  U(x) = e^{-i\varphi(x)\tau^{3}/2} e^{-i\theta(x)\tau^{2}/2}.
\end{equation} This diagonalization is determined only up to a rotation
around the direction of $\Phi$, i.e a right multiplication of $U$ with
$\exp[i\omega(x)\tau^{3}]$. In analogy with the Abelian case (cf.~Eq.~(\ref{Hilu})), we introduce the variables
\begin{equation}
  \label{gaiv} B_{\mu} = U^{\dagger}( A_{\mu}
+\frac{1}{ig}\partial_{\mu})U.
\end{equation} Under a gauge transformation $\Omega$, the new set of
variables transforms as
$$\left[B,\rho,U\right] \stackrel{\Omega}{\rightarrow }
\left[B,\rho,\Omega U\right].$$  In terms of these variables, the
Georgi-Glashow Lagrangian is written as
\begin{equation}
  \label{gegl2}
 {\cal L}\left[B,\rho\right]=-\fract{1}{4}F_{\mu\nu}[B]
F^{\mu\nu}[B]+\fract{1}{2}\left(\partial_{\mu}\rho\partial^{\mu}\rho
-g^{2}\rho^{2}(B^{1}_{\mu}B^{1\,\mu}+B_{\mu}^{2}B^{2\,\mu})\right)
-V(\rho)
\end{equation} and the generating functional is given by
\begin{eqnarray} Z[J,k] &=& \int d\left[B,\rho,U\right]\,\delta
\left[f({}^{U}\!B, \fract{1}{2} \rho \,U\,\tau^{3} U^{\dagger})\right]
\Delta_{FP}[\,\,{}^{U}\!B, \fract{1}{2} \rho \,U\,\tau^{3}
U^{\dagger}]\nonumber \\ & &\quad\exp\left\{iS\left[B,\rho \right]+i\int d^{4}
x \Bigl(\,{}^{U}\!J^{\mu}B_{\mu}+{}^{U}\!k\,\rho
\fract{1}{2}\tau^{3}-\frac{i}{g} J^{\mu} U\partial_{\mu}
U^{\dagger}\Bigr)\right\}
\label{gglgi}
\end{eqnarray} with
$${}^{U}\!B_{\mu} = U( B_{\mu}
+\frac{1}{ig}\partial_{\mu})U^{\dagger},\quad {}^{U}\!J = U^{\dagger} J
U,\quad {}^{U}\!k = U^{\dagger} k U .$$ Gauge transformations affect the
variable $U$ appearing in the measure of the generating functional; they
do not transform the variables appearing in the action. On the other hand,
the Lagrangian still exhibits a local $U(1)$ symmetry
\begin{equation}
  \label{u1}
  B_{\mu} \rightarrow e^{i\omega(x)\tau_{3}}(
B_{\mu}+\frac{1}{ig}\partial_{\mu}) e^{-i\omega(x)\tau_{3}}
\end{equation} reflecting the ambiguity in the definition of $U$. However
if the gauge condition $f(A,\Phi)$ specifies the gauge completely, the
measure will not be invariant. 

This formulation also exhibits the
particular role of the unitary gauge for describing the physics in the Higgs
phase. A complete  gauge fixing to the unitary gauge is achieved with the
following gauge conditions:
\begin{displaymath}
   \Phi-\fract{1}{2}\rho\,V_{0}\tau^{3}V_{0}^{\dagger}=0, \qquad
\tilde{f}[\mathop{\mathrm tr} (\hat{\Phi}A)]=0.
\end{displaymath} The $x$-independent unitary matrix $V_{0}$
eliminates the functional integration over the angular fields specifying
$U$ and
$\tilde{f}$ can be chosen to fix the local $U(1)$ symmetry (Eq.~(\ref{u1})).
Under the change of variables to B (and using the
$x$-independence of $V_{0}$)
$$
\tilde{f}[\mathop{\mathrm tr}(\hat{\Phi}A_{\mu})] \rightarrow
\tilde{f}[B^{3}_{\mu}].
$$ 
In generalizing our studies of the previous
paragraph we discuss in this section the dynamics in the Higgs phase
using a ``radiation gauge''
$$
f[\,A\,] = 0. 
$$ Whenever the gauge condition is of this form,  i.e.,
independent of the Higgs field, all the ambiguities
(\cite{Gribov78,tHooft81}) and obstructions (\cite{Singer78}) which are
encountered when eliminating gauge degrees of freedom in  pure
Yang-Mills theories, and which have been invoked as possible sources for
confinement must also be present in the Georgi-Glashow model. Unlike
the global gauge fixing of the unitary gauge, radiation gauges cannot be
expected to be globally implemented. Horizons associated with  the gauge
conditions and corresponding restrictions to fundamental domains must
be present when implementing the Lorentz or Coulomb gauge and
Abelian monopoles must exist also in the Higgs phase when a
diagonalization gauge is used \cite{tHooft81}. On the other hand, our
calculation indicates that  such obstructions in implementing the
Coulomb gauge should not be relevant for the dynamics in the Higgs
phase. In the representation (\ref{gglgi}) of
$Z$ in terms of gauge invariant variables, the essential argument can be
made quite easily. To this end we treat the gauge condition, for instance
the Lorentz-gauge condition
$$f(\,{}^{U}\!B, \fract{1}{2} \rho \,U\,\tau^{3} U^{\dagger}) =
\partial^{\mu} \Bigl[U( B_{\mu}
+\frac{1}{ig}\partial_{\mu})U^{\dagger}\Bigr] $$ approximately and
neglect in $f$ the contributions from the charged gauge fields
$B^{1,2}_{\mu}$. In the ``gauge invariant'' formulation (\ref{gegl2}) these
two components are massive and on scales significantly larger than
$1/g\rho_{0}$ fluctuations of these fields should be negligible as
compared to the contributions from
$B^{3}_{\mu}$. The approximate gauge condition
$$\partial^{\mu} \Bigl[ U(\fract{1}{2} B_{\mu}^{3}\tau^{3}
+\frac{1}{ig}\partial_{\mu})U^{\dagger}\Bigr] =0 $$ is satisfied
(cf.~Eq.~(\ref{para})) by
$$\partial^{\mu} B_{\mu}^{3} = 0,\quad  \varphi = \mbox{const.}, \quad
\theta = \mbox{const.}\, ,$$  and therefore
$$\int d[U]\delta\Bigl\{\partial^{\mu} [U( B_{\mu}
+\frac{1}{ig}\partial_{\mu})U^{\dagger}]\Bigr\} \approx
\delta(\partial^{\mu} B_{\mu}^{3}) .$$\ The damping of the quantum
fluctuations in the Higgs phase by the mass converts the gauge condition
effectively into an Abelian one. The Gribov horizon cannot be reached
when the charged components acquire a sufficiently large mass. In terms
of Gribov's approximate description (\cite{Gribov78}), the zero-point
fluctuations which reach the Gribov horizon get increasingly damped with
increasing vector boson mass and effectively the horizon disappears when
$$g\rho_{0} \approx 2\Lambda e^{-3\pi^{2}/2g^{2}}$$ with $\Lambda$
regularizing the gauge field fluctuations. This suppression of particular
nonperturbative field configurations in the generating functional as a
consequence of the mass term is operative also in other gauges. In the
Weyl gauge, for instance,
$A_{0}=0$, which can be treated in similar approximate way as the
Lorentz gauge, the topologically nontrivial pure gauges
\begin{equation}  A_{i}=\frac{1}{ig}h^{\dagger}\partial_{i}h, \quad
h=\exp\Bigl(in\pi\,\frac{{\bf x}\cdot\mbox{\boldmath$\tau$}}{\sqrt{{\bf x}^2 +
\xi^2}}\Bigr) 
\label{puga}
\end{equation} are  important for the dynamics of Yang-Mills theories.
Their degeneracy with $A=0$ is lifted when coupled to the Higgs field. For
a spatially constant Higgs field, for instance, the mass term in the gauge
fields gives rise to
 an increase in the energy
$$ \delta E = \frac{n^{2}\pi^{4}}{3}\rho_{0}\left(6 +\,
  _{1}\!F_{2}(1;2,4; -n^2\pi^2)\right) $$ which in turn prevents the
associated nonperturbative phenomenon of vacuum tunneling to occur.
Thus in these gauges, the Higgs phase can be reached perturbatively and
nonperturbative dynamics possibly related to confinement gets
suppressed by coupling  gauge and Higgs fields. \par A different class of
gauges which has been proposed (\cite{tHooft81,Kronfeld87}) to be of
relevance for the description of confinement are the diagonalization
gauges in which the gauge condition is of the form
$$f = E(x)-\fract{1}{2} e(x)\tau^{3}$$ where $E(x)$ is the adjoint quantity
to be diagonalized with $e(x)$ its modulus. The gauge condition becomes
ill defined whenever $E$ vanishes resulting in a monopole singularity of
the gauge field. Formally this gauge condition can be explicitly
implemented in the representation (\ref{gglgi}) of the generating
functional
\begin{eqnarray} Z[J,k] &=& \int d[e]\int d\left[B,\rho,U\right]\,\delta
\left[U\,E\,U^{\dagger}-\fract{1}{2}\, e \, \tau^{3}\right]
\prod_{x}e^{2}(x)\nonumber \\ & &\exp\Bigl\{iS [B,\rho
 ]+i\int d^4 x \,\Bigl({}^U \!\! J^{\mu}B_{\mu}+{}^{U}\!k\,\rho
\fract{1}{2}\tau^{3}-\frac{i}{g} J^{\mu} U\partial_{\mu}
U^{\dagger}\Bigr)\Bigr\}.
\label{gglap}
\end{eqnarray} Unless $E$ vanishes the two angular fields in the $U$
integration can be identified with the angular fields characterizing the
orientation of $E(x)$. Examples which have been discussed are
$$E(x)= F_{12}(x)$$ or the Polyakov gauge
(cf.~\cite{Kronfeld87,Reinhardt97,Jahn98}) with (cf.~Eq.~(\ref{FE3}))
$$E(x)= \frac{1}{2i}
\left(P_{3}(x_{\perp})-P_{3}^{\dagger}(x_{\perp})\right).$$  Both of these
choices have the property that the gauge condition fails in the
perturbative regime. Both gauges are singular for $A=0$ and therefore
gauge fields fluctuating  with small amplitude around zero generate a
large number of monopoles. In such a situation, a small amplitude
expansion of the action will in general not be justified. While also after
gauge fixing the field strength for such configurations remains finite and
actually small, this small value is obtained by a cancellation of the
singular Abelian and the singular commutator term in the field strength.
In these gauges, dropping the commutator term (and possibly neglecting
all but the diagonal gauge field components in a subsequent ``Abelian
projection'' \cite{Suzuki90}) explicitly suppresses these weak field
configurations exponentially beyond the geometric suppression by the
Faddeev-Popov determinant (the $e^{2}(x)$ in the integrand (\ref{gglap}))
by assigning  large values of the action to these field configurations. If
such a description is indeed appropriate for the confining phase of the
Yang-Mills theory, the Higgs phase appears in such gauges to be
separated by nonperturbative dynamics such as the annihilation of the
``Abelian projected'' monopoles.

\section{Conclusion}\setcounter{equation}{0}
 When characterizing phases of
systems that contain gauge degrees of freedom, one is invariably confronted
with the difficulty of disentangling symmetry properties of physical degrees
of freedom from  consequences of the redundancy in the set of variables
in a  locally gauge invariant description. At 
first glance, the standard procedure of representing each gauge orbit by
exactly one representative gauge field appears to resolve the issue.
However,  this resolution is  incomplete and in some instances
incorrect. It is incomplete, since this procedure does not address the
existence of symmetries associated with the gauge degrees of freedom
after elimination of the redundancy. It is incorrect for systems like
pure Yang-Mills theories with their ability to spontaneously break a
discrete part of the gauge symmetry, the center symmetry. 

We have
addressed these issues in the framework of the Abelian and non-Abelian
Higgs models and it is useful to briefly compare and contrast the two
cases. In the Abelian Higgs model, a state that is symmetric under
displacements requires superposition of states of different matter field
winding number. In the Coulomb phase, the copious zeros in $\Phi$ allow
the winding number to change so that the displacement symmetry is
present and gives rise to a  massless photon.  In the Higgs phase,
however, the large magnitude of $\Phi$ freezes the winding number,
prevents displacement symmetry, and thus prevents massless particles. 
Formally, the degrees of freedom relevant to displacement symmetry
disappear in the Higgs phase, and when expressed in terms of gauge
invariant variables, the action displays no residual symmetries.  In the
absence of residual symmetries, there was no need to further consider the
structure of the ground state.  

Unlike the Abelian case, in the Georgi-Glashow model the apparent
displacement symmetry can be continuously connected to unity.  In the
confining phase, the copious zeros in $\Phi$ allow would-be displacements
to be connected to unity by gauge fields which do not
accumulate large contributions to the action.  Thus, unlike in the Coulomb
phase of the Abelian model, where the zeros in $\Phi$ facilitated
displacement symmetry by allowing winding number change, in the
non-Abelian case, the zeros prevent displacement symmetry by allowing
the displacements to  be generated by true gauge transformations.  In the
Higgs phase, however, the finite value of  $\Phi$ prevents displacements
from being connected to unity by gauge fields with low actions,  producing
a residual displacement symmetry and thus a massless particle. When one
transforms to gauge invariant variables in the Higgs phase, the residual
symmetries are explicit.  Given the presence of residual symmetries, we
also determined the structure of the ground states.  In the confined phase,
confinement requires center symmetry so that $\chi$ is symmetrically
distributed around $\pi$ and charge conjugation symmetry is broken.  In
the Higgs phase, we determined that the ground state is charge
conjugation symmetric and breaks center symmetry.  We have not studied
the ground state of the deconfined phase where center symmetry must
be broken, and thus do not know its behavior under charge conjugation.

Ultimately, one would like to understand what happens to aspects of
confinement and other nonperturbative behavior as one goes from the
confined phase to the Higgs phase in the non-Abelian Higgs model.  As a 
starting point, our formal developments have highlighted a number of
similarities between the Abelian transition from the Coulomb to the Higgs
phase and the non-Abelian transition from the confining to the Higgs
phase. 
In the Abelian Higgs model, the
transition from the Higgs to the Coulomb phase is accompanied by a
``condensation'' of 2-dimensional manifolds of zeroes of the Higgs field
which in the unitary gauge can be interpreted as a condensation of
vortices. We similarly expect the corresponding Higgs to confining
phase transition to be accompanied by condensation of the world-lines of
the zeroes of the Higgs field.  In both the Abelian and non-Abelian cases,  a
differential operator appears whose inverse has to develop a gap in the
Higgs phase and the transition to the Higgs phase is accompanied by a
transition from compact to noncompact variables. However, in the
transition to the confining phase additional mechanisms must be also
operative, and we have several hints from this work. An effective field
theory description based on ``radiation'' type gauges indicates that when
decreasing the mass of the vector-particles,  Gribov horizons appear and
the density of vacuum-like states increases near the ground state energy.
Thus, below some sufficiently low value of the Higgs field, one
expects both confinement and nonperturbative phenomena like
instantons. 

\subsubsection*{Acknowledgments}

It is a pleasure for J.W.N. and L. O'R. to acknowledge support by Alexander
von Humboldt Foundation Research Awards and to express their
appreciation for the warm hospitality of the Institute for Theoretical
Physics III at the University of Erlangen, where most of this work was
performed.  F.L. is grateful for the hospitality of the Center for Theoretical
Physics, MIT, where this work was completed.
This work was supported in part by the U.S. Department of
Energy (DOE) under cooperative research agreement
\#DE-FC02-94ER40818 and by the Bundesministerium f\"ur Bildung,
Wissenschaft,  Forschung und Technologie.

\section*{Appendix I: Spontaneous Symmetry Breakdown}
\setcounter{equation}{0}
\renewcommand{\theequation}{I.\arabic{equation}}

In this appendix, we collect some facts about spontaneous symmetry
breaking in nongauged Higgs models which are useful for better
understanding the bulk of this paper. Our main focus will be on the global
phases of the scalar fields, their canonical treatment, the role of boundary
conditions, and a comparison with the standard interpretation of
spontaneous symmetry breaking.

For simplicity let us suppose that the Higgs field $\phi_a(x)$ is  real and
in the fundamental (vector) representation of SO(N), so  that there is only
one fundamental group-invariant, namely
$\rho=\sqrt{\phi_a\phi_a}\geq 0$. The generalization to other cases is
straightforward. The Higgs Lagrangian is
\begin{equation} L=\int \fract12(\partial\phi)^2-V(\rho)  \ .
\end{equation} Since, apart from
$\rho$, the potential contains only constant  parameters, the point
$\rho_0$ where it reaches its minimum must be a  constant. In general
$\rho_0=0$ and $\rho_0\not=0$ correspond to  different phases. Let us
suppose that we are in the phase
$\rho_0\not=0$. Then we are permitted to use the polar  decomposition
$\phi=\rho\hat\phi(\theta)$ with $\hat\phi^2=1$, where the
$\theta$'s are polar angles, and the Lagrangian density may be  written as
\begin{equation} L=L_\rho+L_{\theta,\rho}\ , \qquad L_\rho=\int
\fract12 (\partial\rho)^2-V(\rho) \ ,
\qquad L_{\theta,\rho} =\int
\fract12 \rho^2g_{ab}(\theta)\partial\theta^a\partial\theta^b  
\end{equation} where $g_{ab}$ is the appropriate positive metric. The
minimal configuration for  the Lagrangian is then evidently
$$\rho=\rho_0\not=0 \ ,   \qquad \partial\theta_a=0 \ .  $$ This shows
that the fields $\rho$ and $\theta$ are on a very  different footing. The
field $\rho$ has its minimal value, which is  sharp, dictated by the
potential and is massive, with mass
$V''(\rho_0)>0$. But, although the minimal values of the $\theta$'s   are
constant, their actual values are left completely undetermined and need
not even be sharp, and they are evidently massless Goldstone fields.
We see also that since for $\rho$  the situation would be the same even if
there were no $\theta$ variables,
$\rho$ is related to phase-transitions rather than symmetry  breaking.
Indeed we shall see that even for $\rho_0\not=0$, with the  concomitant
existence of massless Goldstone modes, the  rigid symmetry is not
necessarily broken.

The symmetry is carried by the $\theta$'s and to investigate the
situation  with respect to these we note  that the
$\theta$-momenta, and the $\theta$ phase-space are
\begin{equation}
\pi_a=\rho^2g_{ab}\dot\theta^b \ , \qquad
[\pi_a(x),\theta^b(y)]=ig^b_a\delta(x-y) 
\nonumber
\end{equation} while the
$\theta$-Hamiltonian and the group generators are
\begin{equation} H_{\theta,\rho}=\fract12\int g_{ab}\bigl(\rho^{-2}
\pi^a\pi^b+\rho^2\nabla\theta^a\cdot\nabla\theta^b\bigr)
\quad \hbox{and} \quad Q=\int_\Omega \pi^\theta_a\delta\theta^a \ ,
\end{equation} respectively. We now extract the zero-modes of $\theta$
and
$\pi_\theta$ with respect to the spacial coordinates according to
\begin{equation}
\theta^a(x,t)=\theta^a_0(t)+\tilde\theta^a(x,t) \ ,
\qquad \int_\Omega \tilde\theta^a=0 \ ,
\qquad \theta^a_0={1\over \Omega}\int_\Omega \theta^a  
\end{equation} and
\begin{equation}
\pi_a(x,t)={1\over\Omega}\pi_a^0(t)+\tilde\pi_a(x,t)\ , \qquad
\int_\Omega \tilde\pi_a=0 \ ,\qquad \pi_a^0=\int_\Omega
\pi_a(x,t)        \ .
\end{equation} Then the phase-space decomposes into
\begin{equation} [\tilde\pi_a(x),\tilde\theta_b(y)]
=i\delta_{ab}\Bigl(\delta(x-y)-\Omega^{-1}\Bigr) \ ,
\qquad [\pi_a^0,\tilde\theta_b]=[\tilde\pi^a,\theta^b_0]=0 \ ,
\qquad [\pi_a^0,\theta_0^b]=i \ .
\end{equation} The normalization has been chosen so that $\theta_0$ and
$\pi^0$ satisfy the conventional quantum mechanical relation with no
factor
$\Omega$.

\paragraph{\em The Hamiltonian.} The group generators decompose
into
\begin{equation} Q=Q_0+\tilde Q \ , \qquad Q_0=\pi^0_a\delta\theta^0_a \
,
\qquad \tilde Q=\int_\Omega  \tilde\pi^a\delta\tilde\theta_a  
\nonumber
\end{equation} and the Hamiltonian becomes
\begin{equation} H_\theta=\Bigl({\Omega_\rho\over \Omega^2}\Bigr)
{g_{ab}\pi^a_0\pi_b^0\over 2} +{\pi^a_0\over
\Omega}\int\rho^{-2}g_{ab}\tilde\pi^b_0+\tilde H
\end{equation} where $\Omega_\rho=\int \rho^{-2}$ and
\begin{equation}
\tilde H=\int_\Omega
:\rho^{-2}g_{ab}\tilde\pi^a\tilde\pi_b+\rho^2\nabla\tilde\theta^a
\cdot\nabla\tilde\theta^b: \,   =\int
(\rho^{-2}\tilde\pi^a+i\sqrt{\nabla}\tilde\theta_a)\rho^2
g_{ab}(\rho^{-2}\tilde\pi^b-i\sqrt{\nabla}\tilde\theta^b)\ .
\end{equation} The tilde-vacuum is defined by
\begin{equation}
(\rho^{-2}\tilde\pi^b-i\sqrt{\nabla}\tilde\theta^b)|\rangle=0 \ .
\nonumber
\end{equation} Both of the tilded variables $\tilde\theta$ and $\tilde\pi$
appear quadratically in the Hamiltonian and hence we can normal order
and define a vacuum in the usual way.  Taking the tilded vacuum
expectation value of the Hamiltonian and using
$\langle\rho\rangle=\rho_0$ we obtain
\begin{equation}
\langle H_\theta\rangle\equiv H_0=\rho_0^2{g_{ab}\pi^a_0\pi^b_0\over
2\Omega}\ , \qquad \langle G\rangle=G_0=\pi_0\delta\theta_0 \ .
\end{equation} Since the number of zero-mode variables is finite, $H_0$ is
just  an  ordinary quantum Hamiltonian. However, since the $\theta_0$ do
not appear,  it is not possible to normal order or define a normalized
ground  state ($H_0$ has an infimum but no minimum). If we allow 
infinite norm states then for finite volume the minimization of
$H_0$ requires that $\pi^\theta_0=0$, in which case $Q_0=0$ and  the
system is in the broken but symmetric phase.  However, in the infinite
volume limit $H_0$ becomes zero and if the system were isolated  it
would make no statement about the ground state
$\theta_0$ configuration. Any state compatible with the group relation
$[Q,\theta_0^a]=i\delta\theta_0^b$ would be permissible, in particular the
symmetric and nonsymmetric extremes  $Q_0=0$ and $\theta_0$ sharp.
What determines the choice of configuration is not the system itself but 
the {\it boundary conditions}. For example, with an interaction of the 
form $\epsilon(\theta\cdot\Phi)^2$, where $\Phi$ represents an external 
heath-bath and $\epsilon$ is the coupling constant (trigger), the result 
depends on the order in which the limits $\Omega\rightarrow
\infty$ and
$\epsilon\rightarrow 0$ are taken. If $\epsilon\rightarrow 0$ first, then 
we reach the broken but symmetric phase $Q_0=0$ but if
$\Omega\rightarrow\infty$ first  then we reach the broken unsymmetric
phase, $\theta_0$ sharp, where the value is  dictated by $\Phi$. This
dependence on the boundary conditions corresponds  to the usual
situation for ferromagnets and is supposed to  correspond to what
happens in the early universe.

Although the choice $\theta_0$ sharp looks fairly innocuous from the 
mathematical, or even quantum mechanical, point of view, it makes a 
profound physical difference in field theory because of the factor
$\Omega$. To see this consider either the quantum mechanical amplitude
$A$ or the average energy $\Delta E$ needed for a change
$\theta_0\rightarrow \theta_0+\Delta\theta_0$ in time $T$, namely
\begin{equation} A=\sqrt{{T\over
\Omega}}e^{-i{\Omega(\Delta\theta_0)^2 \over 2T}}
\quad \hbox{and} \quad \Delta E={\Omega\over 2T}(\Delta \theta_0)^2
\ ,      \nonumber
\end{equation} respectively. The critical parameter in both cases is
$\Omega/T$. It is clear that when $\Omega/T$ becomes infinite the
probability  of changing to another value of $\theta_0$, or of changing to
a  symmetric configuration, is zero, and that the reason for this is  that it
would require an infinite energy-input. Thus, once the system  has been
forced into a configuration where $\theta_0$ is sharp by the  boundary
conditions, it is forced to stay in that configuration. The  only question is
when the limit $\Omega/T\rightarrow \infty$ occurs.  It is usually
assumed to occur in two and three space dimensions on  the dimensional
grounds that $\Omega=T^d$ where $d$ is the space dimension. For one
space dimension (two-dimensional space-time),  on the other hand,
$\Omega=T$ so the ratio is not automatically  infinite for infinite
$\Omega$. In this case, whether the symmetry is spontaneously broken
or not depends on the model.

For the path integral, the nonsymmetric case corresponds to inserting  a
delta function  $\delta(\theta_0-\Theta_0)$ in the measure
$d\theta_0$ for the variable $\theta_0$. This corresponds to taking a
single point in the group-orbit of $\theta_0^a$ but this is  natural because
the $\theta_0$ themselves  and, in the  infinite volume limit, their
conjugates $\pi_0$, do not appear in the  Hamiltonian. Furthermore, the
different sectors correspond to the same  physics and are separated by
infinite energy barriers. In  particular, the group transformations that
link the different  sectors are forbidden.

\paragraph{\em  Comparison with Standard Group Statements.} Let us
digress for a moment to consider the conventional spontaneous symmetry
breaking statements  about the generators in the spontaneously broken
case, namely
\begin{equation}
  \label{lo1} [Q,\phi^a(x)]=i\delta\phi^a(x)\qquad \Rightarrow \qquad
\langle \phi_0^a\rangle\not=0\quad\rightarrow\quad
\langle Q^2\rangle\not=0
\end{equation} which links spontaneous breakdown with noninvariance
of the vacuum, and
\begin{equation}\label{lo2}
\langle Q^2\rangle=\int dx\,dy\, \langle j(x)j(y)\rangle=
\Omega\int dx\,\langle j(x)j(0)\rangle
\end{equation} 
where $j(x)$ is the time-component of a local current,
which purports to  show that $Q$ becomes infinite when
$\Omega\rightarrow \infty$.

The generator for the fluctuating part 
$\tilde Q$ of $Q$ is quadratic and can be  normal-ordered in the usual way. It
is therefore zero on the vacuum,  and we then see that these conventional
statements are actually  statements about the zero mode parts, namely
\begin{equation}
\rho_0\not=0 \quad\rightarrow \quad
[Q_0,\theta_0^a]=i\delta\theta^a_0\not=0\quad \rightarrow
\quad Q_0\not=0
\end{equation} and
\begin{equation}
\langle Q^2\rangle=Q_0^2=(\pi_a^0\delta\theta^a_0)^2
\end{equation} From these equations we see that the conventional
conclusion that  the generators become infinite is not correct. What
happens in  (\ref{lo2}) is that the $j$'s are of order $\Omega^{-1}$ and
thus the  whole expression is of order 1. The correct statement seems to
be  that the generators operate in the usual manner on the  phase-space
$\{\pi^0,\theta_0\}$, but, as we have seen,  because of the factor
$\Omega$ the dynamics of this phase-space  becomes trivial in the
infinite $\Omega$ limit.

\section*{Appendix II: Ward Identities of the Displacement Symmetry}
\setcounter{equation}{0}
\renewcommand{\theequation}{II.\arabic{equation}} Quantum
mechanically one identifies the massless photons as Goldstone bosons
associated with the symmetry breakdown of the displacement symmetry
with the help of the associated Ward-Identities. For their derivation we
consider soft modulations of the rigid displacements (\ref{dps})
\begin{equation}
  \label{ldp}
  \Phi(x) \rightarrow e^{-id_{\mu}(x)x^{\mu}}\Phi(x),\quad A_{\mu}(x)
  \rightarrow A_{\mu}(x)+ \frac{1}{e} d_{\mu}(x) .
\end{equation} This is not a gauge transformation; we can restrict the
arbitrary $d_{\mu}(x)$ to not change at all the chosen gauge. In
Coulomb gauge, the following choice
\begin{equation}
  \label{sldp}
  d_{0}(x)=0\, ,\quad \mbox{div}\,{\bf d}(x) = 0
\end{equation} guarantees that only physical variables -- the transverse
gauge fields -are affected by the transformation. These transformations
are transverse to the gauge orbits.  They  change the value of the action
and for infinitesimal transformations
\begin{equation}
  \label{ds}
\delta S = \int d^{4} x d_{\nu}(x)\,\partial_{\mu}
\left[\frac{1}{e} F^{\mu\nu}-x^{\nu}j^{\mu}\right]  
\end{equation} where the current of the charged matter field is given by
$$j_{\mu} = i\,\Phi^{\star}\stackrel{\leftrightarrow}{\partial}_{\mu}\Phi
-2 e A_{\mu} \Phi^{\star}\Phi .$$  Accounting also for the corresponding
changes in the external source terms we obtain the  fundamental
functional identity
\begin{eqnarray}
  \label{fid}
  0 &=& \int d[A, \Phi]\,\delta [\mbox{div} \, {\bf A} ]
e^{iS\left[A,\Phi\right]}e^{i\int d^{4} x\,
  \left(J^{\mu}A_{\mu}+ k^{\star}\Phi+k\Phi^{\star}\right)}
 \int d^{4} x\, d_{\nu}(x)\nonumber\\ &&\qquad {}\cdot \Bigl( \partial_{\mu}
   \Bigl[\frac{1}{e} F^{\mu\nu}-x^{\nu}j^{\mu}\Bigr]
+\frac{1}{e}J^{\nu}-ix^{\nu} ( k^{\star}\Phi-k\Phi^{\star})\Bigr)
\end{eqnarray} which is the quantum mechanical version of 
Eq.~(\ref{maxw}). This identifies the four conserved Noether
currents associated with the displacement symmetry
$$ C^{\mu\nu}=F^{\mu\nu}-ex^{\nu}j^{\mu}$$ 
or, in the case of Coulomb gauge the
two components after  transverse projection.

 Conservation of
the Noether currents $C$ has been derived
 independently of the
 conservation of the charge current $j$.  In order to account for the
 latter we proceed as usual. Under the transformation
$$ \Phi(x) \rightarrow e^{i\alpha(x)}\Phi(x),\quad A_{\mu}(x)
  \rightarrow A_{\mu}(x)
$$ the action changes for infinitesimal $\alpha$
\begin{equation}
  \label{dsp}
\delta S = \int d^{4} x \,\alpha(x)\partial_{\mu}j^{\mu}
\end{equation} and yields
\begin{equation}
  \label{fip}
  0 = \int d[A, \Phi]\,\delta [\mbox{div} \, {\bf A} ]
e^{iS\left[A,\Phi\right]}e^{i\int d^{4}x\, 
  \left(J^{\mu}A_{\mu}+ k^{\star}\Phi+k\Phi^{\star}\right)}
 \int d^{4} x\, \alpha(x)\left( \partial_{\mu}j^{\mu} +i
  k^{\star}\Phi-k\Phi^{\star}\right). \nonumber 
\end{equation} This equation can be used to eliminate the explicit
$x$-dependence in Eq.~(\ref{fid}) by  choosing  $\alpha = d\, x$, with the
transverse but otherwise arbitrary  vector field ${\bf d}$
(cf.~Eq.~(\ref{sldp})) with the result
\begin{equation}
  \label{fi2}
 0 = \int d[A, \Phi]\,\delta [\mbox{div} \, {\bf A} ]
e^{iS\left[A,\Phi\right]}e^{i\int d^{4} x\,
  \left(J^{\mu}A_{\mu}+k^{\star}\Phi+k\Phi^{\star}\right)}\left( \Box {\bf
A}(x)- e
  {\bf j}_{tr}(x) + {\bf J}_{tr}(x)\right)
\end{equation} where $tr$ denotes the transverse component of the
corresponding vectors. This is the quantum mechanical equation of
motion from which  the
standard Schwinger-Dyson equation may be obtained by functional
differentiation with respect to the source $J(y)$ evaluated at $J = 0$
$$ \Box_{x} \langle 0|T( A_{tr}^{i}(x) A_{tr}^{j}(y))|0\rangle +e \langle 0|T(
j_{tr}^{i}(x)
  A_{tr}^{j}(y))|0\rangle +i
\Bigl(\delta_{i,j}-\frac{\nabla_{i}\nabla_{j}}{\Delta}\Bigr)\delta(x-y)=0.$$  Ward
identities always follow from properties of the equations of motion; they
are obtained by applying a particular subset  of variable transformations
which yield the equations of motion. In the case of displacements, this
particular subset happens to coincide with the most general
transformations.

\bibliographystyle{unsrt}

\end{document}